\documentclass[12pt]{article}
\usepackage{comment}
\usepackage{amsmath}
\usepackage{amssymb}
\usepackage{graphicx}
\usepackage{here}
\usepackage{mathtools}
\usepackage{subcaption}
\usepackage{url}
\usepackage{mathtools}
\usepackage[sort&compress, numbers, merge]{natbib}
\usepackage{physics}
\usepackage{todonotes}

\usepackage[colorlinks=true, linkcolor=blue, citecolor=blue, urlcolor=black]{hyperref}

\begin{document}

\def\ps{\mathbf{p}}
\def\PS{\mathbf{P}}

\baselineskip 0.6cm

\def\simgt{\mathrel{\lower2.5pt\vbox{\lineskip=0pt\baselineskip=0pt
           \hbox{$>$}\hbox{$\sim$}}}}
\def\simlt{\mathrel{\lower2.5pt\vbox{\lineskip=0pt\baselineskip=0pt
           \hbox{$<$}\hbox{$\sim$}}}}
\def\simprop{\mathrel{\lower3.0pt\vbox{\lineskip=1.0pt\baselineskip=0pt
             \hbox{$\propto$}\hbox{$\sim$}}}}
\def\tr{\mathop{\rm tr}}
\def\SU{\mathop{\rm SU}}

\begin{titlepage}

\begin{flushright}
OU-HET-1104
\end{flushright}
~

\vskip 1.5cm

\begin{center}

{\Large \bf
Is light thermal scalar dark matter possible?}

\vskip 2cm
{\large
Tomoya Hara, Shinya Kanemura, and Taisuke Katayose
}

\vskip 1.0cm
{\it
Department of Physics, Osaka University, Toyonaka 560-0043, Japan
}

\vskip 3.5cm
\abstract{We study a light thermal scalar dark matter (DM) model with a light scalar mediator mixed with the standard model Higgs boson, including both the theoretical bounds and the current experimental constraints. The thermal scalar DM with the mass below a few GeV is usually strongly constrained by the observation of CMB and/or indirect detection experiments because the leading annihilation mode is S-wave. However, we find that two parameter regions still remain, which are the resonant annihilation region and the forbidden annihilation region. For the both cases, higher partial waves dominantly contribute to the annihilation at the freeze-out era, and the constraint from the cosmological observation is weaker. We consider typical cases of these regions quantitatively, mainly focusing on the mixing angle and the mass of the new particles. Finally, we also discuss the testability of this model at future experiments.}

\end{center}

\end{titlepage}

\tableofcontents
\newpage

\section{Introduction}
\label{sec: Introduction}
 The existence of dark matter (DM) is well established by many cosmological observations, and the amount of DM is accurately determined as $\Omega h^2 = 0.120\pm 0.001$ by the observation of CMB at the Planck experiment\,\cite{Planck:2018vyg}. The thermal DM is one of the most motivated DM candidates which has been studied in many new physics models, and their phenomenological properties have been well studied. Especially when the DM has a weak charge, the DM with the mass around the electroweak scale is favored to explain the present relic abundance, and such a DM is searched by the direct detection experiments as XENON 1T\,\cite{XENON:2018voc} or LUX\,\cite{LUX:2016ggv} very effectively. On the other hand, a relatively light thermal DM particle is difficult to be detected by direct detection experiments, and we also need to consider the other interaction than the weak interaction by the Lee-Weinberg bound\,\cite{Lee:1977ua} which restricts the weak-charged DM particle with the mass lower than a few GeV. 

 The simplest model is a singlet scalar extension of the standard model (SM) with an unbroken $Z_2$ symmetry\,\cite{Silveira:1985rk,McDonald:1993ex,Burgess:2000yq,Barger:2007im,Lerner:2009xg,GAMBIT:2017gge,Kanemura:2010sh,Cline:2013gha}, and only the resonant annihilation region of the Higgs boson and the large mass region still survive. Another possibility is to introduce a mediator particle which makes a connection between DM and SM particles. In this paper we consider a SM singlet scalar particle as the simplest mediator. With this mediator, we focus on the possibilities of the light singlet scalar DM which has not been studied well. A light singlet fermion DM with such a mediator has been discussed in Ref.\,\cite{Dolan:2014ska,Krnjaic:2015mbs, Matsumoto:2018acr, Bondarenko:2019vrb}. The model with a real scalar singlet DM and a real scalar singlet mediator is nothing but a two-scalar-singlet extension\footnote{The complex scalar singlet extension is equivalent to real two-scalar-singlet extension with fewer parameters, and such a model have been studied in Ref.\cite{Coito:2021fgo}. } of the SM with an unbroken $Z_2$ symmetry to make the DM stable,
which is discussed in Ref.\,\cite{Abada:2011qb,Abada:2012hf, Ahriche:2013vqa, Arhrib:2018eex, Maniatis:2020ois, Basak:2021tnj}. In these studies, they suppose that there is another $Z_2$ symmetry called $Z^{\prime}_2$ under which the mediator particle is odd and other particles are even, and the mediator field has a vacuum expectation value after the symmetry breaking. We remove this $Z^{\prime}_2$ symmetry to consider the possibility of the two-scalar-singlet model more generally at the small mass region. The mediator particle has a coupling with the SM Higgs boson and this coupling causes the mixing between them. The interaction between the SM particles and the DM particle is mainly through this mixing, and we study this model by focusing on the mixing angle and masses of the DM and the mediator. The light scalar DM is, however, strongly constrained by the CMB observation because its dominant annihilation occurs in S-wave. We find that there are still surviving parameter spaces, which are the resonant annihilation region where the mediator mass is almost twice as the DM mass and the forbidden annihilation region\footnote{The forbidden annihilation DM with vector mediators have been studied in Refs.\cite{DAgnolo:2015ujb,Okada:2019sbb}.} where the mediator is slightly heavier than the DM\,\cite{Griest:1990kh}. 

This paper is organized as follows. In Section\,\ref{sec: model}, we define the Lagrangian of the two-scalar-singlet extension of the SM model and also the masses and couplings of the physical states. In Section\,\ref{sec: theoretical}, we discuss the constraint from vacuum stability, perturbative unitarity and relic abundance. In Section\,\ref{sec: experimental constraint}, we consider experimental constraints from collider experiments, beam dump experiments, direct detection experiments and cosmological observations. In Section\,\ref{sec: analysis}, we show the results separating into two regions, the resonant annihilation region and the forbidden annihilation region. In Section\,\ref{sec: future prospect}, we discuss the possibility to explore the surviving parameter regions by future experiments. Finally, in Section\,\ref{sec: conclusion}, we summarize the light dark matter model with two-scalar-singlet extension.

\section{The model}
\label{sec: model}
We introduce a gauge-singlet real scalar 
DM and a gauge singlet real scalar mediator in addition to the SM. The Lagrangian is as follows:
\begin{eqnarray}
\label{eq: lagrangian}
\mathcal{L} &=& \mathcal{L}_\mathrm{SM} + \mathcal{L}_\mathrm{DM}+  \frac{1}{2}(\partial_\mu S)^2 +|\partial_\mu \Phi|^2 -  V_{\mathrm{med}}(S,\Phi) \, ,  \\
V_{\mathrm{med}}(S,\Phi) &=&\mu^3_1 S + \frac{m^2_S}{2} S^2 + \frac{\mu_3}{3!} S^3 + \frac{\lambda_S}{4!}S^4  \nonumber\\ 
&& - \mu^2_\Phi |\Phi|^2  + \lambda_\Phi |\Phi|^4  \nonumber \\
&& + \mu_{S\Phi} S |\Phi|^2 + \frac{\lambda_{S\Phi}}{2} S^2 |\Phi|^2 \, ,
\end{eqnarray}
where $\mathcal{L}_\mathrm{SM}$ is the SM Lagrangian without the scalar potential, $\mathcal{L}_\mathrm{DM}$ is the part including DM field $\eta$, and $V(S,\Phi)$ is the scalar potential written in terms of the gauge singlet real scalar $S$ and the $SU(2)_L$ doublet scalar field $\Phi$.

 The DM $\eta$ is stabilized by an unbroken $Z_2$ symmetry under which the DM field is odd and the other fields are even, and $\mathcal{L}_\mathrm{DM}$ can be written as follows: 
\begin{eqnarray}
\mathcal{L}_\mathrm{DM} &=&  \frac{1}{2}(\partial_\mu \eta)^2 - V_\mathrm{DM} (\eta, S, \Phi) \, , \\
V_\mathrm{DM}(\eta, S, \Phi) &=& \frac{m^2_\eta}{2}\eta^2 + \frac{\lambda_\eta}{4!}\eta^4 + \frac{\mu_{\eta S}}{2} \eta^2 S + \frac{\lambda_{\eta S}}{4} \eta^2 S^2 + \frac{\lambda_{\eta \Phi}}{2} \eta^2 |\Phi|^2 \, .
\end{eqnarray}



\subsection{Definition of the physical mediator field}
The vacuum is defined as the minimum of the potential, and physical fields are defined as the expansion of the fields around the vacuum.  
The mediator field $S$ and $SU(2)_L$ doublet field $\Phi$ have non zero vacuum expectation values, and we redefine the scalar fields $H^\prime$ and $h^\prime$ as
\begin{eqnarray}
S &=& H^\prime + v_S \, ,\\
\Phi &=& \left(
\begin{array}{c}
0  \\
(h^\prime + v)/\sqrt{2}  \\
\end{array}
\right) \, ,
\end{eqnarray}
where $v_S$ and $v$ are the vacuum expectation values of $S$ and $\Phi$, respectively. Using the redundancy of the model accompanying the global shift of $S$ by a constant, we can set as $v_S =0$ without loss of generality. This corresponds to fixing $\mu^3_1$ as
\begin{eqnarray}
\mu^3_1 = -\frac{\mu_{S\Phi} v^2}{2} \, ,
\label{eq: mu_1}
\end{eqnarray}
where $v = \sqrt{\mu^2_\Phi/\lambda_\Phi}$ .
Under this condition, the quadratic terms of $H^\prime$ and $h^\prime$ are diagonalized using the mass eigenstates $H$ and $h$ as 
\begin{eqnarray}
V_\mathrm{med} &\supset& \frac{1}{2} (H^\prime, h^\prime) \left( \begin{array}{cc}
m^2_{H^\prime} & m^2_{H^\prime h^\prime} \, , \\
m^2_{H^\prime h^\prime} &  m^2_{h^\prime} \, , \\
\end{array}
\right) \left(
\begin{array}{c}
H^\prime  \\
h^\prime  \\
\end{array}
\right) \\
&=& \frac{1}{2} (H, h) \left( \begin{array}{cc}
m^2_{H} & 0 \\
0 &  m^2_{h} \\
\end{array}
\right) \left(
\begin{array}{c}
H  \\
h  \\
\end{array}
\right) \, ,
\end{eqnarray}
where $ m^2_{H^\prime} = m^2_S$, $m^2_{H^\prime h^\prime} = v \mu_{S \Phi}$, $m^2_{h^\prime} = 2\lambda_\Phi v^2$, and 
\begin{align}
m_H^2 &= \frac{m^2_{H^\prime}+m^2_{h^\prime}-\sqrt{m^4_{H^\prime}+m^4_{h^\prime}+4m^4_{H^\prime h^\prime}-2m^2_{H^\prime}m^2_{h^\prime}}}{2} \, ,\\
m_h^2 &= \frac{m^2_{H^\prime}+m^2_{h^\prime}+\sqrt{m^4_{H^\prime}+m^4_{h^\prime}+4m^4_{H^\prime h^\prime}-2m^2_{H^\prime}m^2_{h^\prime}}}{2}\, , \\
&\left(\begin{array}{c}
 H \\
 h 
\end{array}\right) =
\left(\begin{array}{cc}
 \cos{\theta} & - \sin{\theta}\\
 \sin{\theta} & \cos{\theta}
\end{array}
\right)
\left(\begin{array}{c}
 H^\prime     \\
 h^\prime
\end{array}
\right) \, ,
\end{align}
where 
\begin{equation}
    \tan{2\theta} = \frac{2 m^2_{H^\prime h^\prime}}{m^2_{h^\prime}-m^2_{H^\prime}} \, .
\end{equation}

\subsection{The couplings of the physical fields} 
The potential term can be written in terms of $H$, $h$ and $\eta$ as 
\begin{eqnarray}
    V(\eta,H,h) &=& \frac{m^2_\eta}{2} \eta^2 + \frac{m^2_H}{2} H^2+ \frac{m^2_h}{2} h^2   \nonumber \\ 
    &+& \frac{\mu_{\eta H}}{2}\eta^2 H + \frac{\mu_{\eta h}}{2}\eta^2 h + \frac{\mu_H}{3!} H^3 +\frac{\mu_{Hh}}{2} H^2 h +\frac{\mu^{\prime}_{Hh}}{2} H h^2+\frac{\mu_h}{3!} h^3\nonumber \\
    &+& \frac{\lambda_\eta}{4!}\eta^4  + \frac{\lambda_{\eta H}}{4} \eta^2 H^2 + \frac{\lambda_{\eta h}}{4} \eta^2 h^2 + \frac{\lambda_{\eta H h}}{2} \eta^2 H h  \nonumber \\
    &+& \frac{\lambda_0}{4!}H^4 + \frac{\lambda_1}{3!}H^3 h +  \frac{\lambda_2}{4}H^2 h^2 +\frac{\lambda_3}{3!}H h^3  +\frac{\lambda_4}{4!} h^4 \, ,
\label{eq: physical coupling}    
\end{eqnarray} 
where the expressions of the coupling constants are listed in Appendix\,\ref{sec: appendix}. Their couplings with SM fermions are 
\begin{eqnarray}
\label{eq: yukawa}
\mathcal{L}_\mathrm{Yukawa} &=& \sum_f (- \frac{m_f}{v} \sin{\theta} \bar{f}{f} H +  \frac{m_f}{v} \cos{\theta} \bar{f}{f} h ) \, ,
\end{eqnarray}
where $f$ is a SM fermion, and $m_f$ is the mass of $f$. 

\subsection{Input parameters}
\label{sec: input parameters}
We have 13 free parameters ($\mu_1$, $m_S$, $\mu_3$, $\lambda_S$, $\mu_\Phi$, $\lambda_\Phi$, $\mu_{S\Phi}$, $\lambda_{S\Phi}$, $m_\eta$, $\lambda_\eta$, $\mu_{\eta S}$, $\lambda_{\eta S}$, $\lambda_{\eta \Phi}$) in the scalar sector of the model. Since we impose the condition in Eq.(\ref{eq: mu_1}), $v=246$\,GeV and $m_h=125$\,GeV as input, the number of free parameters are reduced to 10, and we choose ($m_\eta$, $m_H$, $\sin{\theta}$, $\mu_{\eta H}$, $\mu_{\eta h}$, $\mu_H$, $\mu_{Hh}$, $\lambda_\eta$, $\lambda_{\eta H}$, $\lambda_0$) as input parameters.

\subsection{Properties of the mediator}
\label{sec: properties of light Higgs}
Since we are interested in the light DM with the mass below a few\,GeV, the mediator $H$ also must be light. We focus on the region where $m_H\sim \mathcal{O}(m_\eta)$ is satisfied. The mediator $H$ couples to the DM directly, and it couples to SM fermions via the mixing angle $\sin{\theta}$. The partial decay width into SM particles can be written as 
\begin{eqnarray}
\Gamma(H \to \mathrm{SMs} ) = \sin^2{\theta}\;\Gamma_{\mathrm{SM}}(h \to \mathrm{SMs})|_{m_h \to m_H} \, ,
\end{eqnarray}
where $\Gamma_{\mathrm{SM}}(h \to \mathrm{SMs})$ is the decay width of the Higgs boson at the SM. Especially, the width of the decay into SM fermions can be written as 
\begin{eqnarray}
\Gamma(H \to f\bar{f} ) = \sin^2{\theta}\; \frac{g_f m^2_f m_H}{8 \pi v^2}\left(1 -\frac{4m^2_f}{m^2_H} \right)^{3/2} \, ,
\end{eqnarray}
where $g_f$ is the inner degrees of freedom of $f$.
The decay width of $H$ into mesons, however, complicatedly depends on $m_H$ around $m_H \sim 1$ GeV. There are also theoretical uncertainties due to the strong coupling of QCD\,\cite{Gunion:1989we, Raby:1988qf, Monin:2018lee, Winkler:2018qyg}. We adopt the result of Ref.\,\cite{Winkler:2018qyg} in our analysis. 

When $m_H > 2 m_\eta$ is satisfied, $H$ also decays into a DM pair, which contributes to an invisible decay. The partial decay width can be written using the couplings in Eq.\,(\ref{eq: physical coupling}) as,
\begin{eqnarray}
\Gamma(H\to \mathrm{inv}) = \Gamma(H\to \eta \eta) = \begin{dcases}
\frac{1}{32\pi} \frac{\mu^2_{\eta H}\sqrt{m_H^2-4m_\eta^2} }{m_H^2}\, , & (m_H > 2m_\eta) \,  \\
0 \, . & (m_H \leq 2m_\eta) 
\end{dcases}
\end{eqnarray}

As the invisible decay width for $m_H > 2 m_\eta$ is neither suppressed by the mixing angle $\sin\theta$ nor the vacuum expectation value $v$, this is usually much larger than the width of visible decays. Therefore, we can consider almost all $H$ decays invisibly for the region where $m_H>2m_\eta$ is satisfied.  This property largely affects the search for the mediator at colliders. We define two parameter regions; 1) the invisible decay region ($m_H > 2 m_\eta $), and 2) the visible decay region ($m_H \leq 2 m_\eta$). We discuss phenomenology of the mediator for these two regions in Section\,\ref{sec: experimental constraint}.

\section{Theoretical bounds}
\label{sec: theoretical}
\subsection{Vacuum stability condition}
To prevent the vacuum from falling into the infinite depth, the potential must be bounded from below. This condition puts the constraint on the quartic coupling constants of the scalar fields, which can be written as\,\cite{Kannike:2012pe}

\begin{gather}
\lambda_\Phi \geq 0\, , \;  \lambda_S \geq 0\, , \;  \lambda_\eta \geq 0\, ,  \;  \nonumber \\  
3 \lambda_{\eta S} +\sqrt{\lambda_S \lambda_{\eta}} \geq 0  \, , \;
\sqrt{\frac{3}{2}} \lambda_{\eta\Phi} + \sqrt{\lambda_\eta \lambda_{\Phi}} \geq 0 \, , \;
\sqrt{\frac{3}{2}} \lambda_{S\Phi} + \sqrt{\lambda_S \lambda_{\Phi}} \geq 0 \, ,\nonumber \\
\begin{multlined}
    \sqrt{\lambda_\Phi \lambda_\eta \lambda_S} +3 \lambda_{\eta S} \sqrt{\lambda_\Phi} + \sqrt{\frac{3}{2}}\lambda_{\eta \Phi}\sqrt{\lambda_S} +\sqrt{\frac{3}{2}}\lambda_{S \Phi}\sqrt{\lambda_\eta} \nonumber \\  + \sqrt{2 \left(3 \lambda_{\eta S}  +\sqrt{\lambda_S \lambda_{\eta}} \right)\left(  \sqrt{\frac{3}{2}} \lambda_{\eta\Phi} + \sqrt{\lambda_\eta \lambda_{\Phi}}\right) \left( \sqrt{\frac{3}{2}} \lambda_{S\Phi} + \sqrt{\lambda_S \lambda_{\Phi}} \right)  } \geq 0 \, .
\end{multlined}
\end{gather}

The local minimum condition at our vacuum can be written as
\begin{eqnarray}
2\lambda_\Phi m_S^2 > \mu^2_{S\Phi} \, .
\end{eqnarray}

The global minimum condition is, however, too complicated to write down the analytical form even at the tree level. We confirm individually that this condition is satisfied for the region where we analyze (c.f. Section\,\ref{sec: analysis}).

\subsection{Perturbative unitarity}
For perturbative calculation to be good approximation, the coupling constants must be relatively small. The concrete estimation of the bound on the coupling constants is given by considering the partial wave unitarity, which leads the condition as $\mathrm{Re}(a_J)<1/2$\,\cite{Lee:1977eg, Marciano:1989ns}, where $a_J$ is the amplitude of the $J$-th partial wave. Considering the S-wave amplitude ($J=0$) at tree level, we obtain the condition as $|a_0|<1/2$.
For the high energy scattering compared to the mass of the scalar particles where the equivalence theorem holds \cite{Cornwall:1974km,Lee:1977eg}, the amplitudes of scalar-scalar to scalar-scalar scatterings or those that include longitudinal mode of weak gauge bosons are dominated by the scalar quartic coupling constants, because contributions from the s,t,u-diagrams are suppressed by the scattering energy. Taking into account the neutral two body states ($\frac{\eta \eta}{\sqrt{2}}$, $\eta H $, $\eta h $, $\eta Z_L $, $\frac{H H}{\sqrt{2}}$, $H h$, $H Z_L$, $\frac{h h}{\sqrt{2}}$, $h Z_L$, $\frac{Z_L Z_L}{\sqrt{2}}$, $W_L^+ W_L^-$) and diagonalizing the matrix of their amplitudes, we obtain
\begin{eqnarray}
|\lambda_{\eta S}|,  |\lambda_{\eta \Phi}|, |\lambda_{S \Phi}|, |\lambda_\Phi|, \frac{|\lambda_\alpha|}{2}, \frac{|\lambda_\beta|}{2}, \frac{|\lambda_\gamma|}{2} < 8\pi \, ,
\end{eqnarray}
where $\lambda_\alpha, \lambda_\beta, \lambda_\gamma$ are the solutions of the cubic equation for $t$, written as
\begin{multline}
 t^3 - (6\lambda_\Phi + \lambda_\eta + \lambda_S  ) t^2 + (6\lambda_\Phi \lambda_S +6\lambda_\Phi \lambda_\eta + \lambda_S \lambda_\eta -4 \lambda_{S\Phi}^2 -4 \lambda_{\eta \Phi}^2 -\lambda_{\eta S} ^2 ) t \\
 +(-6\lambda_\Phi \lambda_S \lambda_\eta + 6\lambda_\Phi \lambda_{\eta S}^2 +4 \lambda_\eta \lambda_{S\Phi}^2 +4 \lambda_S \lambda_{\eta \Phi}^2 -8 \lambda_{\eta S} \lambda_{\eta \Phi} \lambda_{S \Phi}) =0 \, .
\end{multline}

\subsection{Relic abundance condition}
We consider the thermal DM scenario where the DM particles were in the thermal equilibrium with SM particles in early Universe. Because of the expansion of the Universe, density of DM decreases, and the DM particles decouple from the thermal bath of SM particles. This is called ``freeze-out" and the relic abundance of DM is determined by this mechanism. The most accurate observation of the relic abundance is given by the Planck collaboration\,\cite{Planck:2018vyg} from the observation of the fluctuation of the CMB, which is
\begin{eqnarray}
\Omega h^2 = 0.120 \pm 0.001 \, .
\end{eqnarray}

\begin{figure}[h]
    \centering
    \includegraphics[width=35mm]{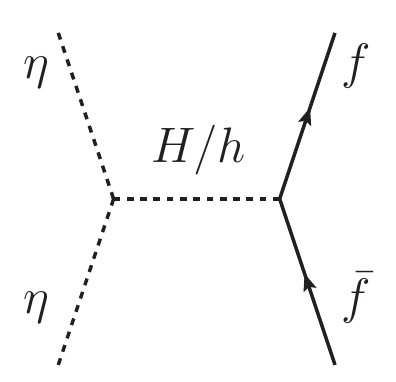}
    \quad
    \includegraphics[width=35mm]{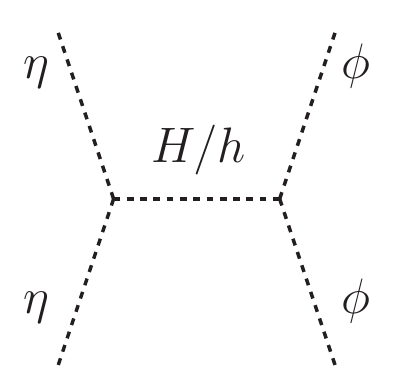}
        \quad
    \includegraphics[width=35mm]{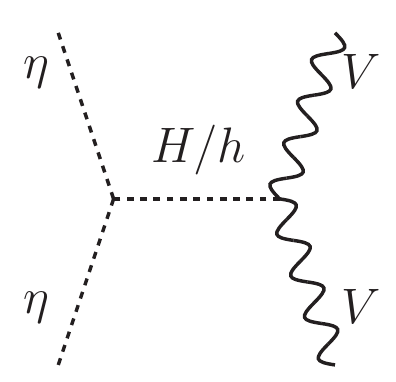}
    \caption{\small \sl Annihilation diagrams of DM via s-channel propagation of the mediator and/or SM like Higgs boson. Here $f$, $\phi$ and $V$ stand for a SM fermion, scalar meson and vector boson respectively.}
    \label{fig: diagrams1}
\end{figure}

\begin{figure}[h]
    \centering
    \includegraphics[width=35mm]{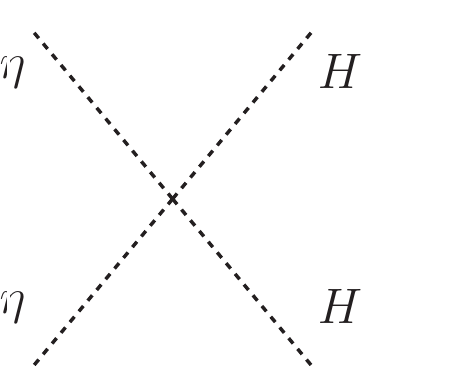}
    \quad
    \includegraphics[width=35mm]{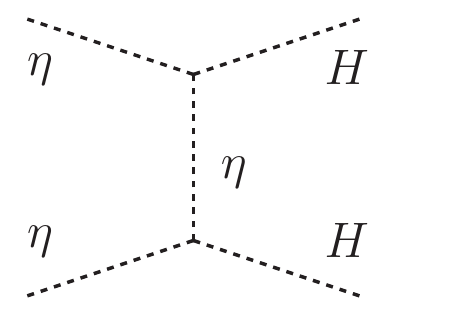}
        \quad
    \includegraphics[width=35mm]{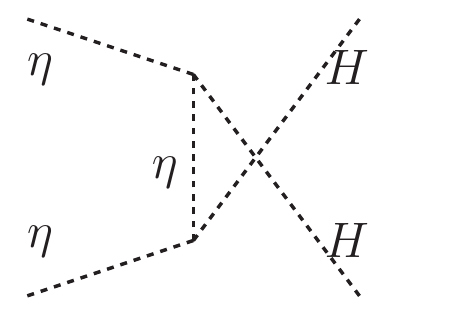}
            \quad
    \includegraphics[width=30mm]{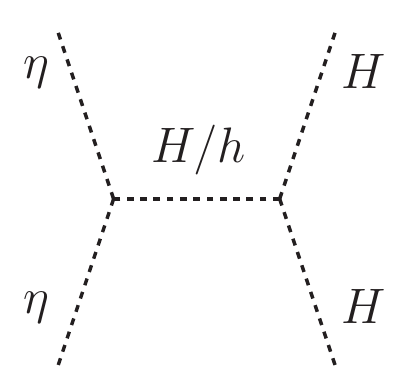}
       \caption{\small \sl Annihilation diagrams of DM into two mediators.}
    \label{fig: diagrams2}
\end{figure}

The annihilation cross section of DM is the most important quantity to calculate the relic abundance.
There are mainly two types of annihilation modes; one is the annihilation into two SM particles through the s-channel propagation of $H$ or $h$, and the other is the annihilation into two mediator particles which finally decay into SM particles. 
Feynman diagrams for each annihilation type are shown in Fig.\,\ref{fig: diagrams1} and Fig.\,\ref{fig: diagrams2}. For the first case, the annihilation cross section is given by

\begin{eqnarray}
\sigma_{\eta\eta \to \mathrm{SMs}} &=& \frac{1}{\sqrt{s-4m_\eta^2}} \left| \frac{-\sin{\theta} \mu_{\eta H}}{s-m_H^2 + im_H \Gamma_H} + \frac{\cos{\theta} \mu_{\eta h}}{s-m_h^2+ im_h \Gamma_h} \right|^2 \Gamma(h \to \mathrm{SMs})|_{m_h \to \sqrt{s}} \, ,
\end{eqnarray}
where $s$ is the square of the energy at the center-of-mass frame.

For the second case, the annihilation cross section is determined by the quartic coupling constant $\lambda_{\eta H}$ and the trilinear coupling constants $\mu_{\eta H}$ and $\mu_{H}$, which can be written as 

\begin{eqnarray}
\sigma_{\eta\eta \to HH} = \frac{\sqrt{s-4m^2_H}}{64\pi^2 s \sqrt{s-4m^2_\eta}} \int d\Omega \left|\lambda_{\eta H} + \frac{\mu^2_{\eta H}}{t-m_\eta^2} + \frac{\mu^2_{\eta H}}{u-m_\eta^2} + \frac{\mu_{\eta H}\mu_H }{s-m^2_H}+\frac{\mu_{\eta h}\mu_{Hh} }{s-m^2_h} \right|^2 \, ,
\label{eq: diagram2}
\end{eqnarray}
where $t=(p_1-p_3)^2$ and  $u=(p_1-p_4)^2$, where $p_1$ and $p_2$ are the in-coming momenta of initial particles, and $p_3$ and $p_4$ are the out-going momenta of final particles. Finally, we solve the Boltzmann equation using the thermal averaged annihilation cross section which can be written as\,\cite{Gondolo:1990dk}
\begin{eqnarray}
\label{eq: sigma v}
\langle \sigma v \rangle &=& \frac{1}{8m_\eta^4 T K^2_2(m_\eta/T)} \int_{4m_\eta^2}^\infty \sigma (s-4m_\eta^2)\sqrt{s}K_1(\sqrt{s}/T)ds \, ,
\end{eqnarray}
where $K_1$ and $K_2$ are the modified Bessel functions of the second kind of order 1 and 2, respectively. 

\section{Experimental constraints}
\label{sec: experimental constraint}

\subsection{Collider experiment}
\label{sec: collider experiment}
DM particles and mediator particles are created via the decay of the SM-like Higgs boson or mesons, and also can be directly produced by collider experiments. In this section, we discuss the constraints from collider experiments for two regions; 1) the invisible decay region and 2) the visible decay region, as we have defined in Section\,\ref{sec: properties of light Higgs}.

\subsubsection*{Direct production of the mediator}
The mediator particle can be directly produced via the process $f\bar{f}\to Z \to Z^* H$. Among the several collider experiments, L3 collaboration at LEP\,\cite{L3:1996ome} give the most stringent constraint on $\sin\theta$ as a function of $m_H$ for the invisible decay region and also for the visible decay region. In Ref.\,\cite{Winkler:2018qyg}, it is pointed out that $H$ is mostly invisible even for the visible decay region because it decays outside of the detector if $m_H < 2m_\mu $ where $m_\mu$ is the muon mass. We refer to their result for the visible decay region.


\subsubsection*{Higgs boson decay}
The SM like Higgs boson $h$ can decay into the mediator or DM via the couplings in Eq.(\ref{eq: physical coupling}). Neglecting the three body decay of $h$, the partial decay widths can be written as 
\begin{eqnarray}
\Gamma(h \to H H) &=&\frac{1}{32\pi} \frac{\mu^{ 2}_{Hh}\sqrt{m_h^2 -4m_H^2}}{m_h^2} \, , \\
\Gamma(h \to \eta \eta) &=&\frac{1}{32\pi} \frac{\mu^{ 2}_{\eta h}\sqrt{m_h^2 -4m_\eta^2}}{m_h^2} \, .
\end{eqnarray}

These partial decay widths can contribute to the Higgs invisible decay width, written as follows: \\ \\
1) Invisible decay region
\begin{eqnarray}
\mathcal{B}(h\to \mathrm{inv.}) &=&  \frac{ \Gamma ( h\to \eta \eta ) + \Gamma (h\to H H)}{\Gamma(h\to \mathrm{SMs}) + \Gamma (h\to \eta \eta ) + \Gamma (h\to H H)} \, ,
\end{eqnarray} \\ \\
2) Visible decay region
\begin{eqnarray}
\mathcal{B}(h\to \mathrm{inv.}) &=&  \frac{ \Gamma ( h\to \eta \eta ) }{\Gamma(h\to \mathrm{SMs}) + \Gamma (h\to \eta \eta )+ \Gamma (h\to H H) } \, ,
\end{eqnarray}
where $\Gamma(h\to \mathrm{SMs}) = 4.1$\,MeV\,\cite{ParticleDataGroup:2020ssz} is the total decay width of the Higgs boson in the SM. The ATLAS collaboration\,\cite{ATLAS:2018bnv} gives a constraint on the Higgs invisible decay as $\mathcal{B}(h\to \mathrm{inv})<0.13$.

\subsubsection*{$\Upsilon$ decay}  
This model induces new $\Upsilon$ decay modes such as $\Upsilon \to  \eta \eta$ or  $\Upsilon \to  H H$, and also induce the decay along with a photon $\Upsilon \to \gamma H$. We focus on the latter process as it gives more severe constraint than the former processes. The branching fraction can be written as\,\cite{Wilczek:1977zn},
\begin{eqnarray}
\frac{\mathcal{B}(\Upsilon \to \gamma H    )}{\mathcal{B}(\Upsilon \to \mu^+ \mu^-)} &=&  \sin^2{\theta}  \frac{G_F m_b^2}{\sqrt{2}\pi \alpha} \sqrt{1-\frac{m_H^2}{m_\Upsilon^2}}\, ,
\end{eqnarray}
where $ \mathcal{B}(\Upsilon \to \mu^+ \mu^-) (= 2.48 \times 10^{-2} $\,\cite{CLEO:2004tkr}) is the branching fraction of $\Upsilon \to \mu^+ \mu^-$, $G_F$ is the Fermi constant, $\alpha$ is the fine structure constant, $m_b$ is the mass of bottom quark, and $m_\Upsilon$ is the mass of $\Upsilon$. \\ \\ 
1) Invisible decay region \\
 The Belle collaboration\,\cite{Belle:2018pzt} and the BaBar collaboration\,\cite{BaBar:2010eww, BaBar:2008aby} give the constraint on the branching fraction of invisible decay as a function of the mass of mediator, which is roughly $\mathcal{B}(\Upsilon \to \gamma +\mathrm{inv.})<10^{-6}\mathchar`-10^{-5}$ for the mediator $H$ with $m_H \sim \mathcal{O}(1)$\,GeV. \\ \\
2) Visible decay region \\
When the mediator only decays into SM particles, we can search for $\Upsilon \to \gamma H$ focusing on the process $\Upsilon \to \gamma + \ell^+ \ell^- $ or $\Upsilon \to \gamma  + \mathrm{jets}$.  The CLEO experiment\,\cite{CLEO:2008jdl} and BaBar experiment\,\cite{BaBar:2012wey,BaBar:2012sau} give constraints on the leptonic decay mode, as $\mathcal{B}(\Upsilon \to \gamma + \mu^+ \mu^-)<10^{-6}\mathchar`- 10^{-5}$ and $\mathcal{B}(\Upsilon \to  \gamma +\tau^+ \tau^-)<10^{-5}\mathchar`-10^{-4}$ for the mediator with $m_H \sim \mathcal{O}(1)$\,GeV.
The BaBar experiment\,\cite{BaBar:2011kau} also searches for the hadronic decay process of the mediator, which gives the constraint $\mathcal{B}(\Upsilon \to \gamma +\mathrm{jets})<10^{-6}\mathchar`-10^{-4}$, and we show the constraint from this process in Fig.\,\ref{fig: forbidden}.

\subsubsection*{$B$ meson decay}
$B$ meson can decay into the mediator $H$ with a $K$ meson when $ m_H < m_B - m_K $ is satisfied, where $m_B$ and $m_K$ are the masses of $B$ meson and $K$ meson, respectively. The decay width of this process can be written as\,\cite{Hiller:2004ii, Dolan:2014ska}
\begin{eqnarray}
\Gamma(B\to K H) &=& \frac{|g_{Hsb}|^2}{16\pi m^3_B}\left(\frac{m^2_B-m^2_K}{m_b-m_s}  \right)^2 f^2_K(m_H) \sqrt{(m^2_B-m^2_K-m^2_H)^2-4m^2_K m^2_H} \, ,  
\end{eqnarray} 
where $g_{Hsb}$ is the effective coefficient of the coupling of $H$, $s$ and $b$ written as 
\begin{eqnarray}
g_{Hsb} &=& \frac{3\sqrt{2} G_F m_b m_t^2 V^*_{ts} V_{tb}}{16\pi^2 v} \sin{\theta} \, ,
\end{eqnarray}
and $f_K(q)$ is the scalar form factor parameterized as\,\cite{Ball:2004ye}
\begin{eqnarray}
f_K(q) &=& \frac{0.33}{1-q^2/37.5\,\mathrm{GeV}^2} \; .
\end{eqnarray}\\
1) Invisible decay region \\
The Belle\,\cite{Belle:2013tnz} and BaBar\,\cite{BaBar:2010oqg} experiments search for $B^+ \to K^+ + \nu \bar{\nu}$ which gives the same signal for $B^+ \to K^+ H$ if $H$ decays invisibly. The upper limit at 90\% confidence level is 
\begin{eqnarray}
\mathcal{B}(B^+ \to K^+ + \mathrm{inv.}) = \frac{\Gamma(B^+ \to K^+ + \mathrm{inv.})}{\Gamma(B^+ \to \mathrm{any})}  < 1.6 \times 10^{-5} \, ,
\end{eqnarray}
where $\Gamma(B^+ \to \mathrm{any}) = 4.0 \times 10^{-13}$\,GeV\,\cite{ParticleDataGroup:2020ssz} is the total decay width of the $B^{+}$ meson. We show the constraint on $\sin\theta$ from this condition in Fig.\,\ref{fig: invisible all}. \\ \\
2) Visible decay region \\
The LHCb\,\cite{LHCb:2015nkv,LHCb:2016awg}, BaBar\,\cite{BaBar:2008jdv} and Belle\,\cite{Belle:2009zue} experiments search for $B^+  \to K^+ + \mu^+ \mu^-$ or $B^0 \to K^* + \mu^+  \mu^-$. This gives the constraint on the leptonic decay of $H$, and we show the result of the LHCb experiment in Fig.\,\ref{fig: forbidden}, which gives the most stringent constraint on this decay mode.
\subsubsection*{$K$ meson decay}
$K$ meson can decay into the mediator $H$ and $\pi$ when $m_H < m_K - m_{\pi}$ is satisfied, where $m_{\pi}$ is the mass of $\pi$. The decay width can be written as\,\cite{Deshpande:2005mb, Marciano:1996wy}  
\begin{eqnarray}
\Gamma(K^\pm \to \pi^\pm H) &=& \frac{|g_{Hds}|^2}{16\pi m^3_K}\left(\frac{m^2_{K^\pm}-m^2_{\pi^\pm}}{m_s-m_d}  \right)^2  \sqrt{(m^2_{K^\pm}-m^2_{\pi^\pm}-m^2_H)^2-4m^2_{\pi^\pm} m^2_H} \, , \\
\Gamma(K_L \to \pi^0 H) &=& \frac{|\mathcal{I}(g_{Hds})|^2}{16\pi m^3_{K_L}}\left(\frac{m^2_{K_L}-m^2_{\pi^0}}{m_s-m_d}  \right)^2  \sqrt{(m^2_{K_L}-m^2_{\pi^0}-m^2_H)^2-4m^2_{\pi^0} m^2_H} \, ,
\end{eqnarray}
where $g_{Hds}$ is the effective coefficient of the coupling of $H$, $s$ and $d$ written as 
\begin{eqnarray}
g_{Hds} &=& \frac{3\sqrt{2} G_F m_s m_t^2 V^*_{ts} V_{td}}{16\pi^2 v} \sin{\theta} \, .
\end{eqnarray}
1) Invisible decay region \\
The NA62 collaboration\,\cite{NA62:2020xlg, NA62:2021zjw,NA62:2020pwi} searches for $K^+ \to \pi^+ + \mathrm{invisible}$, and they report 
\begin{eqnarray}
\mathcal{B}(K^+ \to \pi^+ + \mathrm{inv.}) = \frac{\Gamma(K^+ \to \pi^+ + \nu \bar{\nu})}{\Gamma(K^+ \to \mathrm{any})} = (10.6^{+4.0}_{-3.4} \pm 0.9 ) \times 10^{-11}  \, ,
\end{eqnarray}
where $\Gamma(K^+ \to \mathrm{any}) (= 5.3 \times 10^{-17}$\,GeV) is the total decay width of the $K$ meson. In our case, the signal can be regarded as $\mathcal{B}(K^+ \to \pi^+  H)+ \mathcal{B}(K^+ \to  \pi^+ + \nu \bar{\nu} )_\mathrm{SM}$, where $\mathcal{B}(K^+ \to \pi^+ + \nu \bar{\nu})_\mathrm{SM} = (8.4 \pm 1.0) \times 10^{-11} $\,\cite{Buras:2015qea}, and we can set the constraint on the mixing angle not to exceed the observed branching fraction. We show the constraint on mixing angle as the function of $m_H$ in Fig.\,\ref{fig: invisible all}, and the structure seen around $m_H=0.1\mathchar`- 0.2$\,GeV is due to the background process $K^+ \to \pi^+ \pi^0 $. \\ \\
2) Visible decay region \\
The NA62 collaboration\,\cite{NA62:2020xlg, NA62:2021zjw,NA62:2020pwi} also gives the constraint on the process $K^+\to \pi^+  H$ for the visible decay region with almost same magnitude as invisible decay region. 

The KTeV collaboration\,\cite{KTeV:2003sls, KTeV:2008nqz} gives the upper limit on the leptonic decay of $K_L$ as
\begin{eqnarray}
\label{eq: kle}
\mathcal{B}(K_L \to \pi^0 + e^+ e^-) &=& \frac{\Gamma(K_L \to \pi^0 + e^+ e^-)}{\Gamma(K_L \to \mathrm{any})} < 2.8 \times 10^{-10}  \; \; (\mathrm{at}\; 90 \% \; C.L.) \, , \\
\mathcal{B}(K_L \to \pi^0 +\mu^+ \mu^-) &=& \frac{\Gamma(K_L \to \pi^0 + \mu^+ \mu^-)}{\Gamma(K_L \to \mathrm{any})} < 3.8 \times 10^{-10}  \; \; (\mathrm{at}\; 90 \% \; C.L.) \, , 
\label{eq: klmu}
\end{eqnarray}
where $\Gamma(K_L \to \mathrm{any}) (= 1.3 \times 10^{-17}$\,GeV\,\cite{ParticleDataGroup:2020ssz}) is the total decay width of $K_L$. Considering that the KTeV experiment is a fixed target experiment, $H$ must decay into leptons promptly enough before the detector. The analysis including such an effect has been done in Ref.\cite{Winkler:2018qyg}, and we refer to their result. 
We show the constraint on $\sin\theta$ from NA62 and KTeV collaboration in Fig.\,\ref{fig: forbidden}.

\subsection{Beam dump experiment}
Beam dump experiments have the sensitivity for the boosted long-lived particle which is created by the collision of accelerated proton or electron beam and a fixed target. The mediator $H$ can be long-lived enough for sub-GeV region, and we can constrain the mixing angle $\sin\theta$ not to generate $H$ excessively. 
The CHARM experiment\,\cite{CHARM:1985anb}, which is performed with 400\,GeV proton beam and copper as a fixed target, reported the null result for the new particle, and gives the most stringent constraint on our model. As the analysis of the CHARM experiment for the light scalar mixed with the SM Higgs boson has been done by several studies\,\cite{Winkler:2018qyg,Clarke:2013aya,Bezrukov:2009yw,Schmidt-Hoberg:2013hba,Alekhin:2015byh}, we refer to Ref.\,\cite{Winkler:2018qyg} and the constraint is shown in Fig.\,\ref{fig: forbidden}.  
The PS191 experiment is the fixed target experiment with the 19.2\,GeV proton beam, which is analyzed in Ref.\cite{Gorbunov:2021ccu} for the light scalar mixed with the SM Higgs boson, and we also show this result in Fig.\,\ref{fig: forbidden}.

\subsection{Direct detection experiment}
\label{sec: direct detection experiment}
Direct detection experiments search for DM in the galactic halo by elastic scatterings with nucleus. Even though we focus on the GeV to sub-GeV DM, there are still strong constraint on the scattering cross section with nucleon. The elastic scattering cross section of DM and nucleon can be written in the non-relativistic limit as 

\begin{eqnarray}
\sigma_{\eta N} = \frac{\mu^2}{4\pi m^2_\eta}\left(\frac{\mu_{\eta H}C_{HN}}{ m^2_H}+\frac{\mu_{\eta h} C_{hN}}{ m^2_h} \right)^2\, ,
\end{eqnarray}

where $\mu = (m_N m_\eta)/(m_N + m_\eta)$ is the reduced mass of $\eta$ and the nucleon $N$, and $C_{HN}$ and $C_{hN}$ are effective couplings of $H$ with $N$ and $h$ with $N$, written as\,\cite{Shifman:1978zn}
\begin{eqnarray}
C_{HN} &=& - \frac{\sin{\theta} m_N }{v} \left(\frac{6}{27}f^{N}_{TG} +\sum_{q=u,d,s} f^N_{Tq}\right)\, , \\
C_{hN} &=&  \frac{\cos{\theta} m_N }{v} \left(\frac{6}{27}f^{N}_{TG} +\sum_{q=u,d,s} f^N_{Tq}\right)\, ,
\end{eqnarray}
where $f^N_{TG}$ and $f^N_{Tq}$ are the nucleon scalar form factors defined as

\begin{eqnarray}
f^{N}_{Tq} &=& \frac{\langle N|m_q \bar{q}{q}  |N\rangle}{m_N}\, , \\
f^{N}_{TG} &=& 1-\sum_{q=u,d,s} f^{N}_{Tq} \, ,
\end{eqnarray}
and they are evaluated as\,\cite{Belanger:2018ccd} 
\begin{gather}
    f^p_{Tu}=0.0153, \quad f^p_{Td}=0.0191,\quad f^p_{Ts}=0.0447, \\
    f^n_{Tu}=0.0110, \quad f^n_{Td}=0.0273,\quad f^n_{Ts}=0.0447. 
\end{gather}
Among several direct detection experiments, the CRESST-III experiment\,\cite{CRESST:2019jnq} and the DarkSide-50 experiment\,\cite{DarkSide:2018bpj} give the most stringent constraints on GeV to sub-GeV DM mass regions. They are shown in Fig.\ref{fig: invisible all}.

\subsection{Constraint from CMB}
\label{sec: CMB constraint}
Even in the era after the freeze-out of DM, some portion of DM annihilate into SM particles and inject energies into primordial plasma. Such a process modifies the thermal history of Big Bang nucleosynthesis, and affects the power spectrum of CMB.    
Fluctuation of CMB is observed with a great accuracy by the Planck experiment\,\cite{Planck:2018vyg}, which strictly constrains other energy injection processes than those in the SM. This gives an upper limit on the S-wave DM annihilation cross section as

\begin{eqnarray}
f_{\mathrm{eff}}\frac{\langle \sigma v \rangle_{\mathrm{s}}}{m_\eta} & < & 3.2 \times 10^{-28} \mathrm{cm^3 s^{-1} GeV^{-1} } \, ,
\end{eqnarray}
where $f_{\mathrm{eff}}$ is the fraction of the energy released by the DM annihilation into the primordial plasma at the redshift of $z\simeq 600$. For our model, $f_{\mathrm{eff}}$ is estimated to be $\mathcal{O}$(0.1)\,\cite{Padmanabhan:2005es}.  To explain the thermal relic abundance, $\langle \sigma v \rangle \simeq 3\times 10^{-26} \mathrm{cm^3 s^{-1}}$ is required at the freeze-out era, which is significantly larger than the upper limit on the S-wave annihilation cross section for GeV to sub-GeV DM masses.  As scalar DM particles  basically annihilate in S-wave, it seems that such a light scalar DM model is already ruled out. There are, however, two exceptional cases where the contribution from higher partial waves become important\,\cite{Griest:1990kh}. The first case is that the mediator mass is slightly larger than the twice of the DM mass, which is called the resonant annihilation region, where a DM pair resonantly annihilates into SM particles via the s-channel propagation of the mediator. The second case is that the mediator is slightly heavier than the DM, which is called the forbidden annihilation region, where a DM pair can annihilates into two mediators only when DM is thermalized enough. We discuss these two cases in Section\,\ref{sec: analysis}.

\subsection{Observation of cosmic rays}
When DM has a large enough annihilation cross section, DM particles in the galaxy annihilate into SM particles and emit a detectable amount of cosmic rays. 
However, when we focus on the resonant annihilation region or the forbidden annihilation region as discussed in Section\,\ref{sec: CMB constraint}, the annihilation cross section is suppressed for the galactic DM, and the constraint from cosmic rays is always satisfied.

\subsection{Constraint from BBN}
The theory of the Big Bang nucleosynthesis is well established and the prediction of the fraction ratio of $^1$H, $^2$H, $^3 $He and $^4$He matches with the cosmological observations\,\cite{Cyburt:2015mya}. This strongly supports the SM, and we can give a constraint on new physical models. If there exist new long lived particles, they can decay into mesons or leptons during the nucleosynthesis. These particles react with the proton (p) or the neutron (n), and change the n/p ratio from that of the SM prediction. For our case, without disturbing BBN, the new scalar particle $H$ must decay rapidly enough\,\cite{Fradette:2017sdd,Kohri:2001jx}, from which the constraint on $\sin\theta$ can be given as a function of $m_H$ in Fig.\,\ref{fig: forbidden}. 

\subsection{Constraint from SN1987A}
In some parameter regions, the new light scalar particle can be probed by astronomical observations, and one of the most stringent constraints comes from the observation of supernovae. In the SM, the gravitational binding energy of a supernova is released only by the emission of neutrinos. The supernova neutrinos from SN1987A were observed for over 10 seconds by the Kamiokande\,\cite{Kamiokande-II:1987idp} and IMB\,\cite{Bionta:1987qt}, and the energy emission rate by neutrinos is calculated as $L_\nu = \mathcal{O}(1)\times 10^{53} $ erg/sec, which is consistent with the total released binding energy $E_\mathrm{tot}< 6\times 10^{53}$ erg\,\cite{Raffelt:1987yt,Turner:1987by}. If there is a new scalar particle, it may contribute to the additional energy emission from the supernova core and make cooling rate too fast. The case with a light scalar particle mixed with the SM like Higgs boson has been studied by Ref.\,\cite{Dicus:1978fp,Krnjaic:2015mbs,Dev:2020eam}, and we can set the constraint on the parameter space by the condition that energy emission from new particles does not exceed the contribution from the neutrino emission. In our case, $H$ is mainly produced by the nucleon bremsstrahlung process $NN\to NNH$ through the mixing angle, and the upper limit on the mixing angle can be set by the condition that this process does not occur so often. On the other hand, a large mixing angle leads to an effective reabsorption process of $H$, and we cannot constrain such regions. We refer to Ref.\,\cite{Dev:2020eam} and the constraint is shown in Fig.\,\ref{fig: forbidden} around the region with $m_H<0.2$\,GeV and $\sin^2{\theta}\sim 10^{-11}\mathchar`- 10^{-9}$.

\section{Analysis}
\label{sec: analysis}
As we mentioned in Section\,\ref{sec: CMB constraint}, we focus on the resonant annihilation region and the forbidden annihilation region for GeV to sub-GeV DM masses. We take some criteria as defined in the following section, and we discuss the surviving parameter region.

\begin{figure}[t]
    \centering
    \includegraphics[width=80mm]{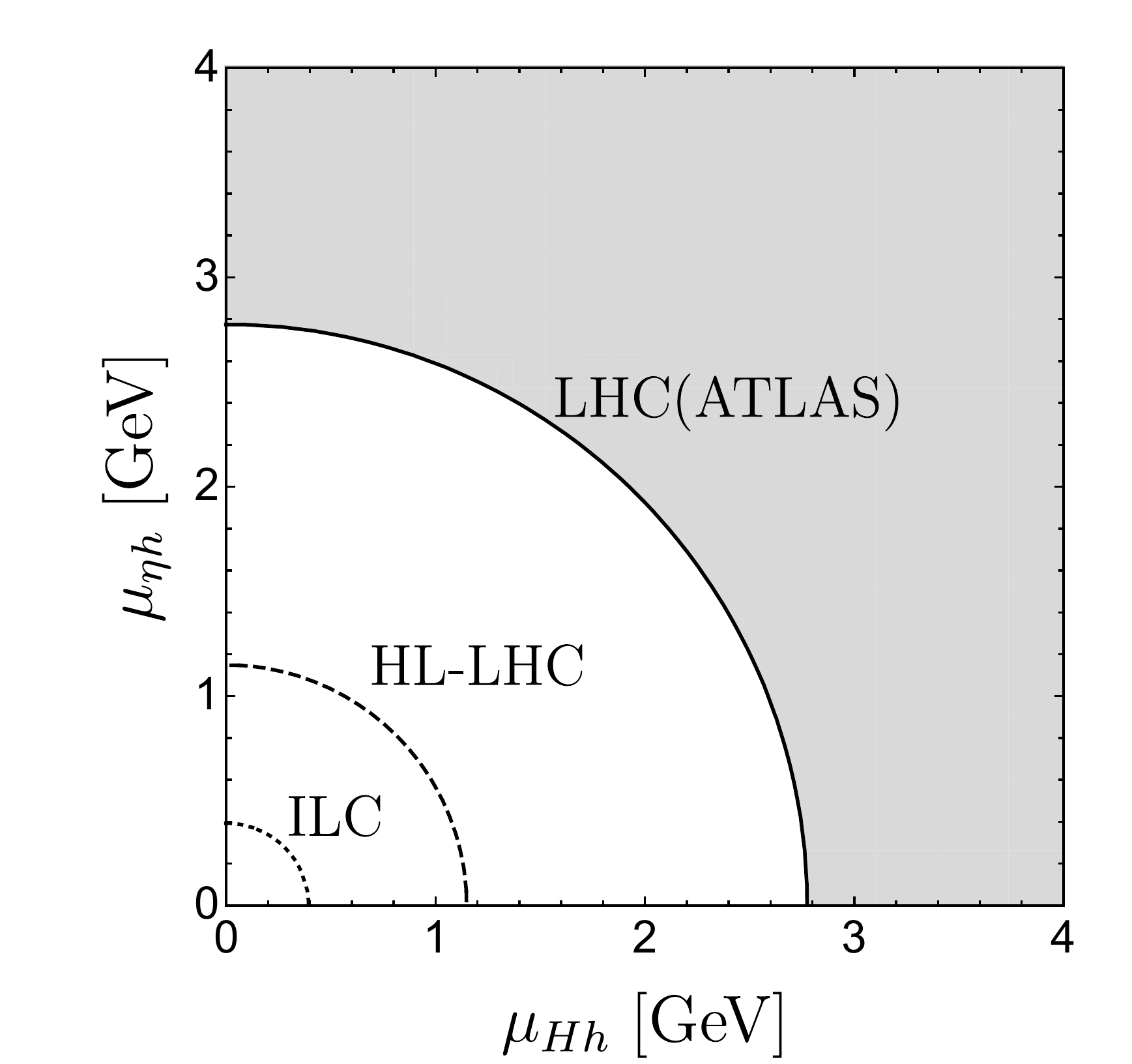}
    \caption{\small \sl The constraint on $\mu_{Hh}$ and $\mu_{\eta h}$ from the Higgs invisible decay results. We also show the projected sensitivity from HL-LHC and ILC. See Sec.\ref{sec: experimental constraint} and Sec.\ref{sec: future prospect} for more details.}
    \label{fig: higgs invisible decay}
\end{figure} 

\subsection{Resonant annihilation region}
\label{sec: resonant region}
For the resonant annihilation region, where $m_H \simeq 2m_\eta$ is satisfied, only the annihilation into SM particles is possible at the freeze-out era. Among the 10 free parameters ($m_\eta$, $m_H$, $\sin{\theta}$, $\mu_{\eta H}$, $\mu_{\eta h}$, $\mu_H$, $\mu_{Hh}$, $\lambda_\eta$, $\lambda_{\eta H}$, $\lambda_0$) mentioned in Section\,\ref{sec: input parameters}, $\mu_H$, $\lambda_\eta$, $\lambda_{\eta H}$ and $\lambda_0$ do not contribute to experimental constraints nor the relic abundance condition. Therefore, we can choose arbitrary values for these parameters under the theoretical constraints, namely, $0<\lambda_\eta, \lambda_{\eta H}, \lambda_0<1$ and $\mu^2_H<3\lambda_0 m^2_H$. The parameter $\mu_{Hh}$ is experimentally constrained only by the Higgs invisible decay results, and the parameter $\mu_{\eta h}$ is constrained by the Higgs invisible decay results and direct detection experiments. The constraint on these parameters from Higgs invisible decay are shown in Fig.\ref{fig: higgs invisible decay}. Here, we only focus on the region $\mu_{\eta h}$, $\mu_{Hh}>0$ to satisfy the vacuum stability condition. 

\begin{figure}[t]
    \centering
    \includegraphics[width=110mm]{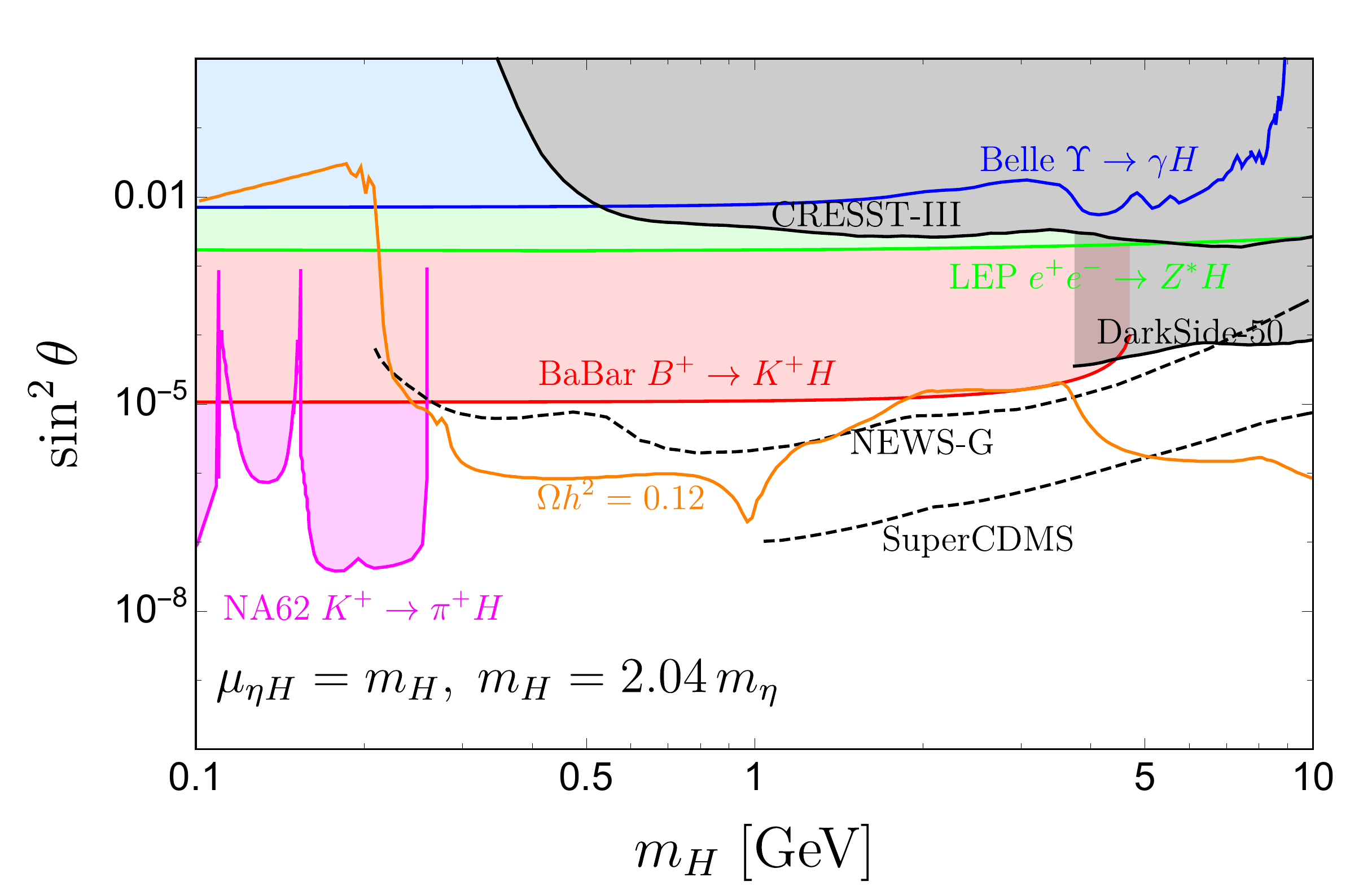}
    \includegraphics[width=110mm]{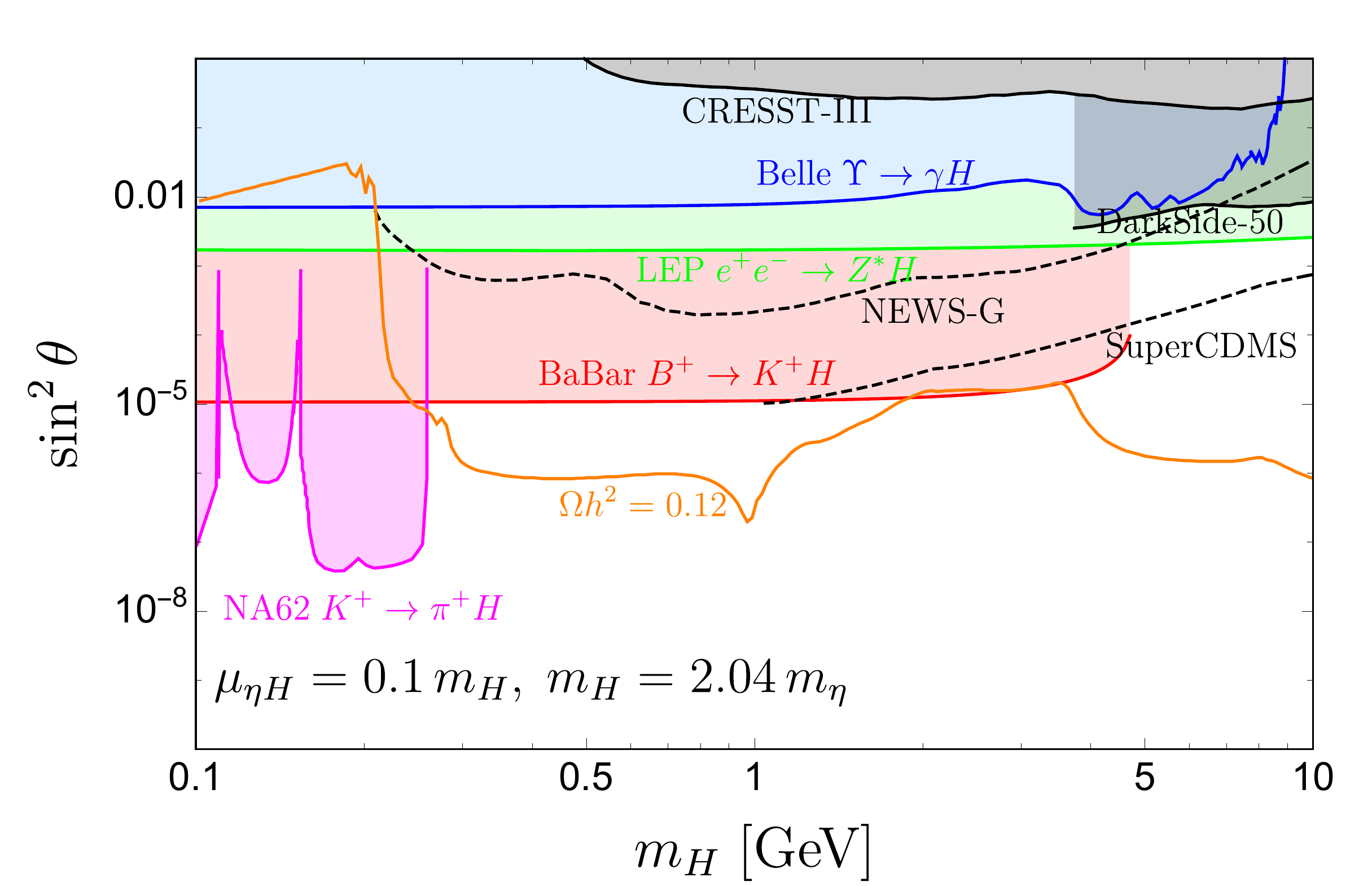}
    \caption{\small \sl The surviving parameter space on the plane of $m_H$ and $\sin^2\theta$ for the resonant annihialtion region. We show the constraints from Belle ($\Upsilon$ decay) as a blue line, LEP (direct production of $H$) as a green line, BaBar ($B^+$ decay) as a red line and NA62 ($K^+$ decay) as a magenta line. Two black solid lines correspond to the constraints from the direct detection experiments, CRESST-III and DarkSide-50 respectively. The two dashed black lines correspond to the prospect of the future direct detection experiments, NEWS-G and SuperCDMS (see Section\,\ref{sec: future prospect}). We also show the contour which satisfies the relic abundance condition as an orange line. The upper figure corresponds to the case with $\mu_{\eta H}=m_H$, $m_H=2.04\,m_\eta$, and the lower figure corresponds to the case with $\mu_{\eta H}=0.1\,m_H$, $m_H=2.04\,m_\eta$.}
    \label{fig: invisible all}
\end{figure} 

\begin{figure}[h]
    \centering
    \includegraphics[width=110mm]{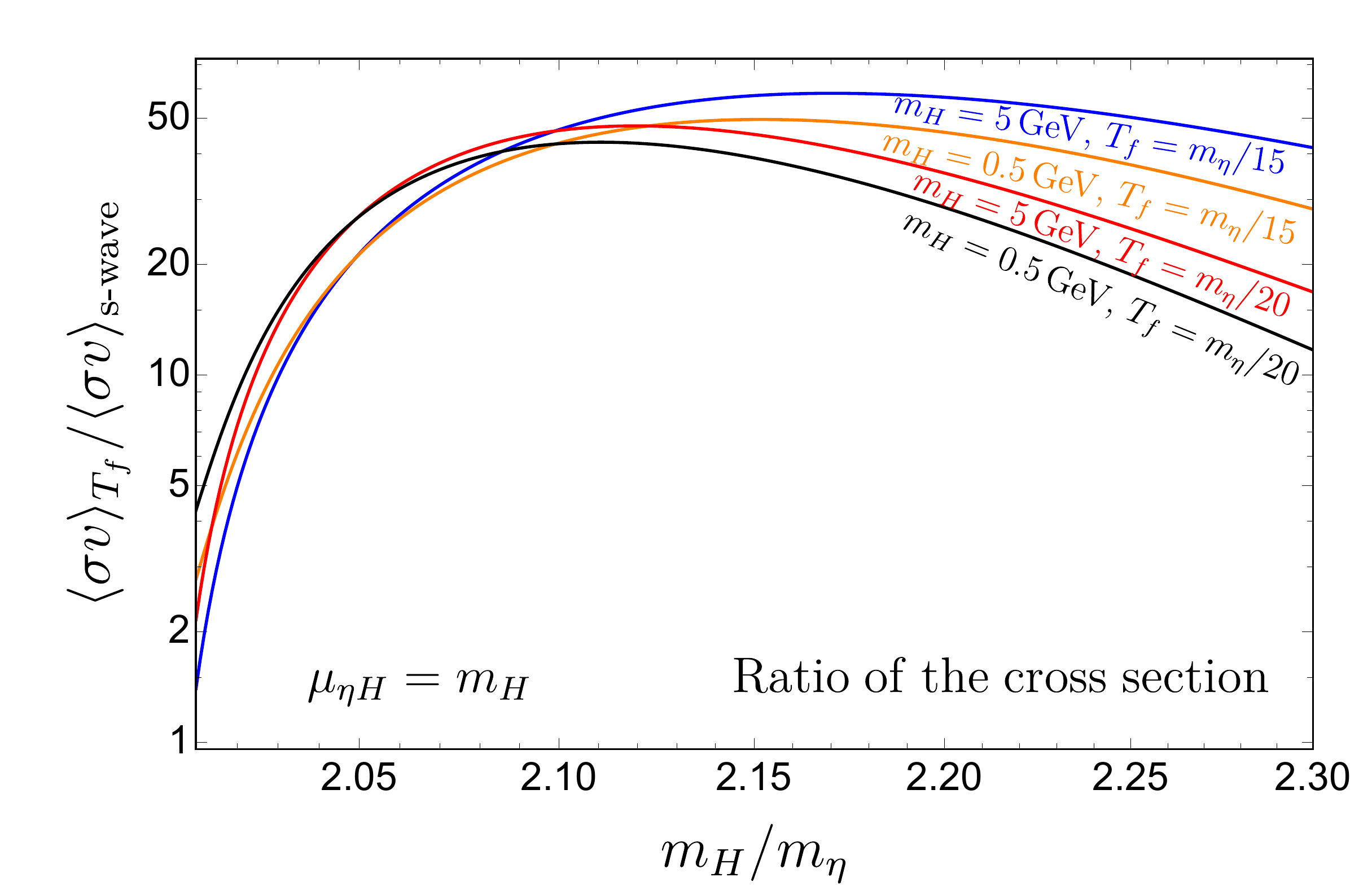}
    \includegraphics[width=110mm]{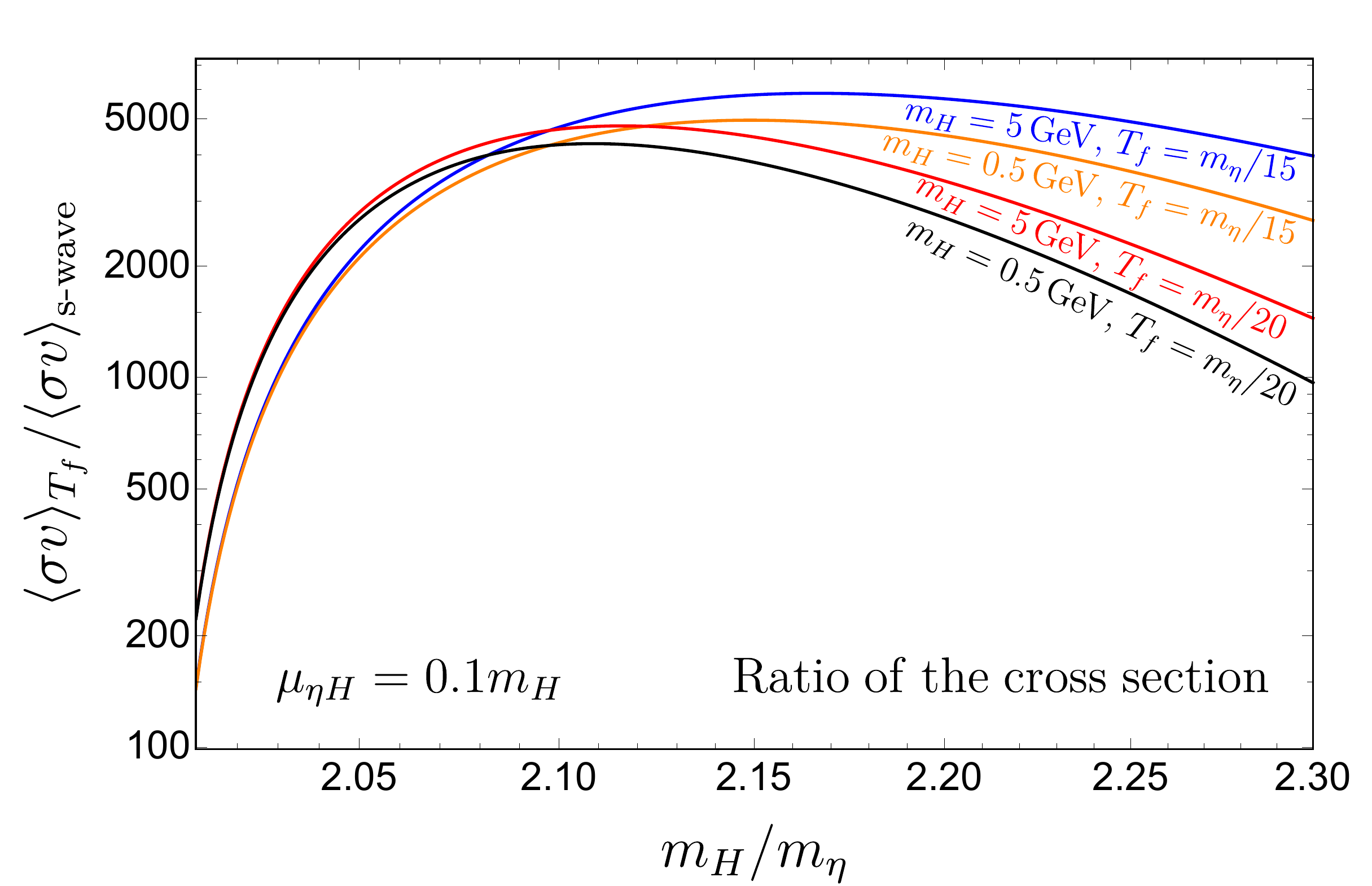}
    \caption{\small \sl The ratio of the annihilation cross section of DM at freeze-out era and that of s-wave. We show the case with $m_H=5,0.5$\,GeV and $T_f = m_\eta /15,20$ for $\mu_{\eta H} =m_H$ (upper figure) and $\mu_{\eta H} =0.1\,m_H$ (lower figure).}
    \label{fig: ratio}
\end{figure} 
The remaining parameters are ($m_\eta$, $m_H$, $\sin{\theta}$, $\mu_{\eta H}$). We have the constraint from relic abundance condition and several experimental constraints on these parameters. We show the surviving parameter space on the plane of the mediator mass $m_H$ and $\sin^2{\theta}$ in Fig.\,\ref{fig: invisible all} for the case with $m_H=2.04\,m_\eta=\mu_{\eta H}$ and $m_H=2.04\,m_\eta=10\,\mu_{\eta H}$. The black shaded region is excluded by the direct detection experiments, whatever value of $\mu_{\eta h}$ is chosen in allowed by the Higgs invisible decay results (c.f. Fig.\ref{fig: higgs invisible decay}). Direct detection experiments can have more sensitivities if the future collider experiments do not observe the Higgs invisible decay. We note that the lines which explain observed relic abundance (orange lines) are almost same for these two cases, and the reason is explained as follows. The annihilation process is regarded as the on-shell production of $H$ and its decay into SM particles, and the production rate of $H$ is proportional to $\mu^2_{\eta H}$ and the branching fraction to SM particles is proportional to $\mu^{-2}_{\eta H}$. Then, the total annihilation cross section of DM to SM particles does not depend on $\mu_{\eta H}$. We also show the ratio of the thermal averaged annihilation cross section at the freeze-out era and that for S-wave in Fig.\ref{fig: ratio}, which justifies our parameter choice regarding the CMB constraint (c.f. Section\,\ref{sec: CMB constraint}). 

\begin{figure}[t]
    \centering
    \includegraphics[width=110mm]{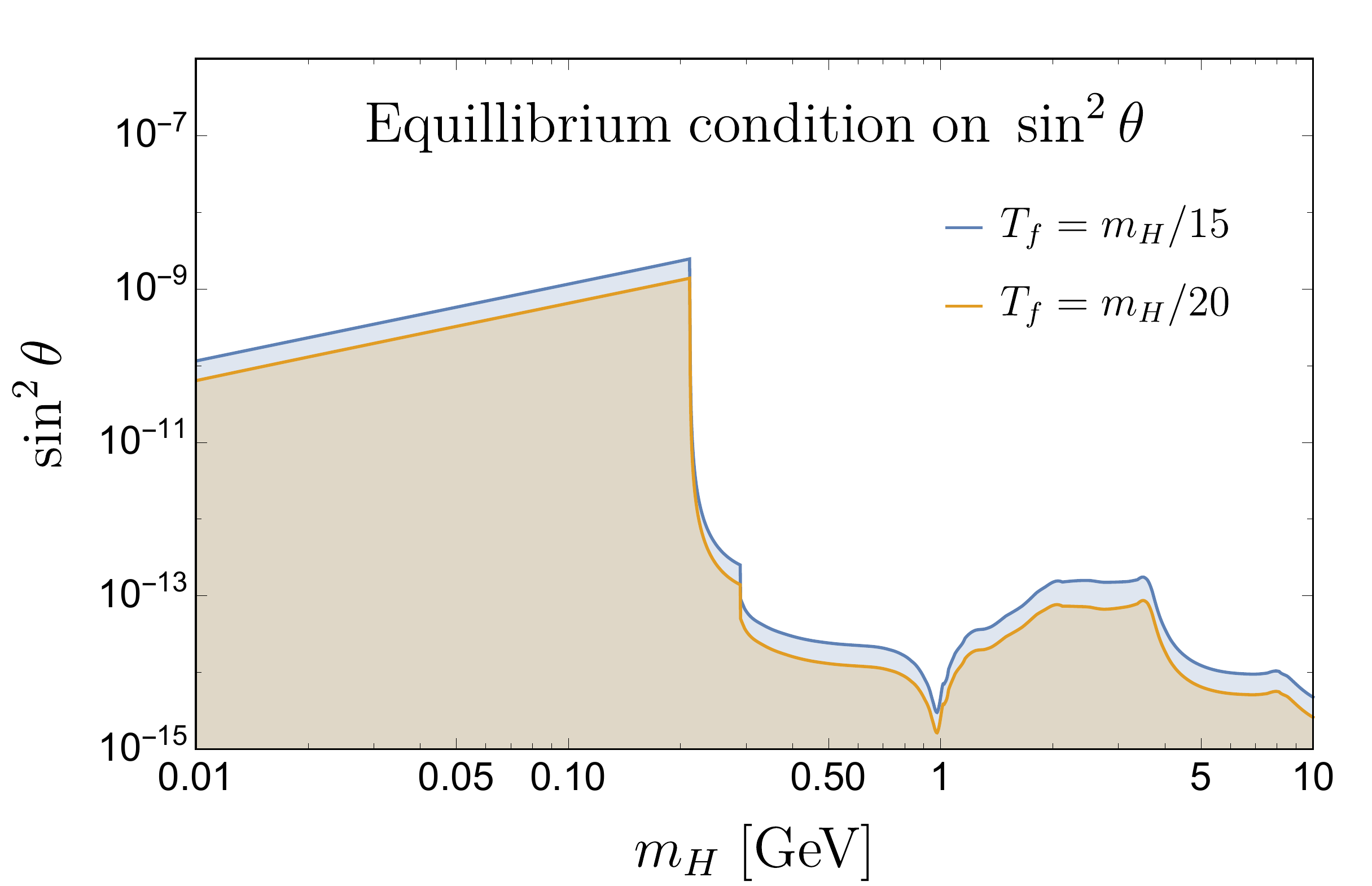}
    \caption{\small \sl The constraint on $\sin^2{\theta}$ from the equilibrium condition of mediator and SM particles. Here we show the plots for two cases: $T_f=m_H/15$ and $T_f=m_H/20$.}
    \label{fig: eq_condition}
\end{figure} 

\subsection{Forbidden annihilation region}
For the forbidden annihilation region, where $m_H \simeq m_\eta$ and $m_H > m_\eta$ are satisfied, the annihilation of DM into two mediators becomes important. In this region, chemical equilibrium of DM and SM particles is maintained through the DM annihilation into two mediator particles and its inverse process as in Fig.\,\ref{fig: diagrams2}. Here we implicitly assume that mediator particles are in equilibrium with SM particles through the decay and inverse decay of the mediator or the process which changes the number of mediator such as $H f \leftrightarrow \gamma f$ or $H \gamma \leftrightarrow f \bar{f}$. To maintain the equilibrium, these processes must occur faster than the expansion of the universe, whose condition can be written as 
\begin{eqnarray}
H(T_f) & < & \Gamma(T_f; \, H\to\mathrm{SMs}) + \sum_{\phi} \Gamma(T_f; \,H+\phi \to \mathrm{SMs}) \, ,
\end{eqnarray}
where $\phi$ is an arbitrary light SM particle, and
\begin{eqnarray}
\Gamma(T_f; \, H\to\mathrm{SMs}) &=& \left\langle \frac{1}{\gamma} \right\rangle \Gamma(H\to\mathrm{SMs}) \, ,\\ 
\Gamma(T_f; \, H+\phi \to  \mathrm{SMs}) &=& n_{\phi} \langle \sigma v \rangle_{H+\phi \to \mathrm{SMs}}|_{T=T_f} \, .
\end{eqnarray}
Here, $\gamma$ is the Lorentz factor which contributes to prolonging the lifetime of $H$, and $n_\phi$ is the number density of $\phi$.  For the forbidden annihilation region, we can consider $T_f\sim m_\eta/20 \sim m_H/20$, and we set $\gamma=1$. Moreover, we can neglect the contribution from $\Gamma(T_f; \, H+\phi \to \mathrm{SMs})$ compared to $\Gamma(T_f; \,H\to\mathrm{SMs})$ because of the factor $n_\phi \propto T^3_f$ and the additional SM coupling constant. Then, it is enough to consider the condition $H(T_f) < \Gamma(H\to \mathrm{SMs})$, and the constraint on the mixing angle $\sin{\theta}$ is shown in Fig.\,\ref{fig: eq_condition}, assuming $T_f = m_H/15\mathchar`-20$.
\begin{figure}[t]
    \centering
    \includegraphics[width=110mm]{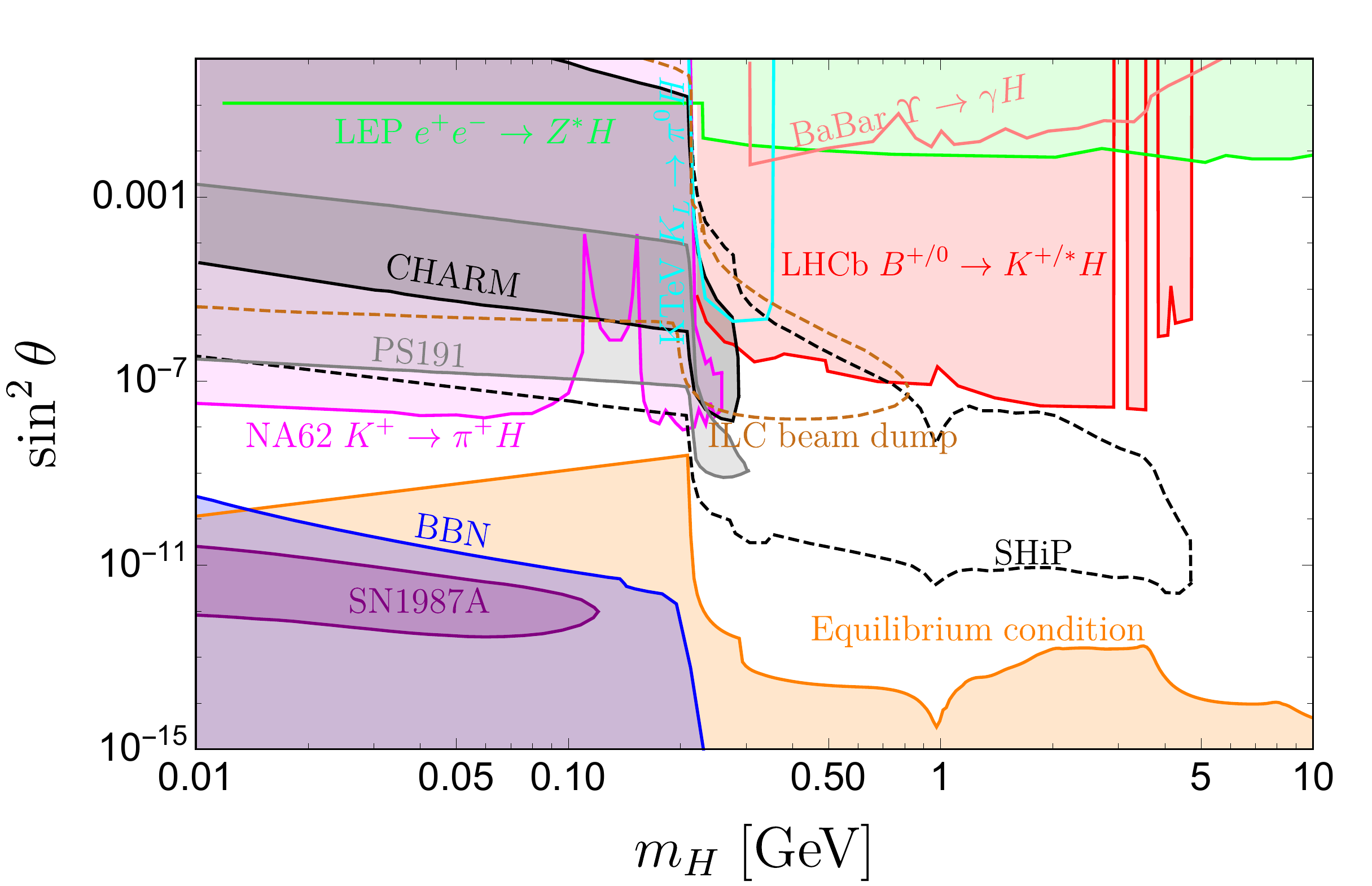}

    \caption{\small \sl The surviving parameter space on the plane of $m_H$ and $\sin^2\theta$ for the forbidden annihilation region. We show the constraints from BaBar ($\Upsilon$ decay) as a pink line, LEP (direct production of $H$) as a green line, LHCb ($B^{+/0}$ decay) as a red line, NA62 ($K^+$ decay) as a magenta line, KTeV ($K_L$ decay) as a cyan line, and CHARM (beam dump) as a black line, BBN as a blue line, the observation of SN1987A as a purple line and the equilibrium condition (at $T_f=m_H/15$) as an orange line. We also show the prospect of future beam dump experiments, SHiP as a black dashed line and ILC beam damp experiment as a brown dashed line. }
    \label{fig: forbidden}
\end{figure} 

\begin{figure}[t]
    \centering
    \includegraphics[width=110mm]{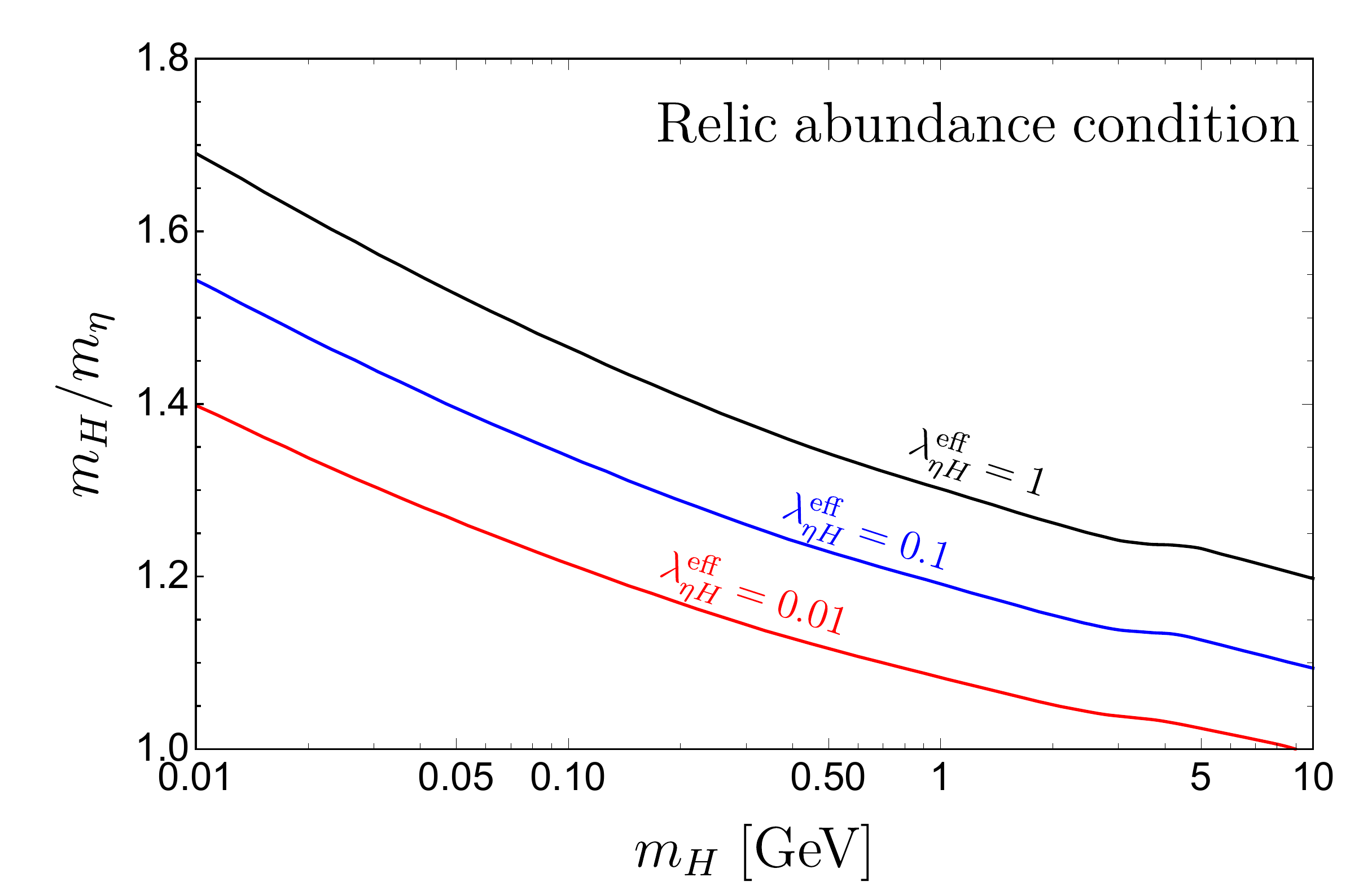}
    \caption{\small \sl The ratio of $m_\eta$ and $m_H$ which satisfy the relic abundance condition assuming $\lambda_{\eta H} =1,\, 0.1, \, 0.01$.}
    \label{fig: 4point}
\end{figure}

Among the 10 free parameters ($m_\eta$, $m_H$, $\sin{\theta}$, $\mu_{\eta H}$, $\mu_{\eta h}$, $\mu_H$, $\mu_{Hh}$, $\lambda_\eta$, $\lambda_{\eta H}$, $\lambda_0$) mentioned in Section\,\ref{sec: input parameters}, $\lambda_\eta$ and $\lambda_0$ do not contribute to experimental constraints or relic abundance condition. Therefore we can choose arbitrary values for these parameters within the theoretical constraints, namely $0<\lambda_\eta, \lambda_0<1$. The parameter $\mu_{\eta h}$ contributes to the Higgs invisible decay and direct detection experiments, while $\mu_{Hh}$ contributes to the Higgs visible decay. However, if $m_H$ and $\sin\theta$ are small enough, $H$ does not decay inside the detector, and the decay mode ($h\to HH$) is invisible. In such a case, we have same constraint as the resonant annihilation region, which is shown in Fig.\ref{fig: higgs invisible decay}. The parameters $m_H$ and $\sin\theta$ are constrained by several experiments or observations, which are shown in Fig.\ref{fig: forbidden}. Here, we do not show the constraint from direct detection experiments unlike the case with the resonant annihilation region, because we can set arbitrary small value for $\mu_{\eta H}$ and  $\mu_{\eta h}$ for forbidden annihilation region. Then, direct detection experiments do not always have sensitivity. The parameters $m_\eta$, $m_H$, $\mu_{\eta H}$, $\mu_H$ and $\lambda_{\eta H}$ are constrained by the relic abundance condition, which are involved in the diagrams at Fig.\ref{fig: diagrams2}. Since the kinetic energies of $\eta$ and $H$ are small enough compared to $m_\eta\simeq m_H$ at freeze-out era, we can set $s=4m^2_H$ and $t=u=0$ in Eq.(\ref{eq: diagram2}). We then obtain the effective four point scalar coupling of $\eta^2 H^2$ as
\begin{eqnarray}
\lambda^{\mathrm{eff}}_{\eta H} &=& \lambda_{\eta H} - 2\,\frac{\mu^2_{\eta H}}{m^2_H} + \frac{\mu_{\eta H} \mu_H}{3 m^2_H} + \frac{\mu_{\eta h} \mu_{Hh}}{4m^2_H-m^2_h}\, .
\end{eqnarray}
We show the contour of the ratio $m_H/m_\eta$ which satisfies the relic abundance condition assuming $\lambda^{\mathrm{eff}}_{\eta H}=1,\,0.1,\,0.01$ in Fig.\,\ref{fig: 4point}.

\section{Future prospect}
\label{sec: future prospect}

Some of the surviving parameter regions of this model will be explored at future
experiments. The constraint from the invisible decay of the SM-like Higgs boson can be improved by the HL-LHC experiment\,\cite{Cepeda:2019klc} and the international linear collider (ILC) experiment\,\cite{Baer:2013cma,Barklow:2017suo}. The projected sensitivities are $\mathcal{B}(h\to \mathrm{inv.})\leq2.5 \%$\,(HL-LHC) and $\mathcal{B}(h\to \mathrm{inv.})\leq0.3 \%$\,(ILC), respectively. The contours correspond to these values are shown in Fig.\ref{fig: higgs invisible decay}. The constraint from $\Upsilon$ and $B$ meson decays shall be improved by the Belle II experiment\,\cite{Aushev:2010bq}, and the constraint from the $K_L$ decay can be improved by the KOTO experiment\,\cite{KOTO:2018dsc}. Future direct detection experiments such as LZ\,\cite{LZ:2015kxe}, SuperCDMS\,\cite{Cushman:2013zza} and NEWS-G\,\cite{news} will have more sensitivity for the region we are focusing on. There is also a planning beam dump experiment SHiP\,\cite{Alekhin:2015byh,Winkler:2018qyg}, which can cover the large parameter space of visible decay region. The ILC beam dump experiment\,\cite{Kanemura:2015cxa,Sakaki:2020mqb} is one option of the ILC experiment, and it also may be able to have a sensitivity for large parameter space. The prospect of the sensitivities at SuperCDMS and NEWS-G are also shown in Fig.\,\ref{fig: invisible all}, and the expected sensitivity at the SHiP and ILC beam dump experiments are shown in Fig.\,\ref{fig: 4point} with dashed lines.

\section{Conclusion}
\label{sec: conclusion}

We have studied the two-scalar-singlet extension of the SM, focusing on the light mass region below a few\,GeV. The CMB observation constrains the S-wave annihilation cross section as $\langle \sigma v \rangle \lesssim 10^{-26} \mathrm{cm}^3 \mathrm{s}^{-1}$ for such a mass region, which conflicts with the required cross section $\langle \sigma v \rangle = 3 \times 10^{-26} \mathrm{cm}^3 \mathrm{s}^{-1}$ to explain the observed relic abundance. However, this is not the case for the resonant annihilation region and the forbidden annihilation region because the main contribution to the annihilation is from higher partial waves. We have studied the surviving parameter space and future prospect of this model for these two regions, by considering theoretical bounds and experimental constraints quantitatively. 

In the resonant annihilation region, the mediator particle is slightly heavier than twice the mass of DM, and the annihilation of DM occurs on the resonance at the s-channel propagation of the mediator. The mediator is mainly decay into the pair of DM, and such a particle can be searched by the meson decay with missing energy. The direct detection experiments also have the sensitivity for this region, and constrain the few GeV DM mass region. There is still surviving parameter region around $m_H=1$\,GeV, and most of the surviving parameter region can be covered by the future collider experiments as Belle II .

In the forbidden annihilation region, the mediator particle is slightly heavier than the DM, and the annihilation of DM is mainly into pair of mediators. The mediator decays into SM particles, and such a particle can be searched by the meson decay along with the charged lepton pair, and also beam dump experiments. For the DM itself, there is a constraint on the ratio between DM and mediator mass, to make the annihilation process $\eta \eta \to H H$ more effective, and this process is independent of mixing angle. We have found that there is large surviving parameter region, and some of this region can be covered by future beam dump experiments such as SHiP and ILC.

\section*{Acknowledgments}

This work was supported, in part, by the Grant-in-Aid on Innovative Areas, the Ministry of Education, Culture, Sports, Science and Technology, No.\,16H06492, and by the JSPS KAKENHI Grant No.\,20H00160.

\appendix
\section{Coupling constants} 
We list here the expression of the coupling constants defined in Eq.(\ref{eq: physical coupling}) from the coupling constants defined in the first Lagrangian Eq.(\ref{eq: lagrangian}) as follows:
\label{sec: appendix}
\begin{eqnarray}
\mu_{\eta H} &=& \mu_{\eta S} \cos{\theta} - \lambda_{\eta \Phi} v \sin{\theta} \, , \\
\mu_{\eta h} &=& \mu_{\eta S} \sin{\theta} + \lambda_{\eta \Phi} v \cos{\theta}\, , \\
\mu_H &=& -3\lambda_{S\Phi}v \cos^2{\theta} \sin{\theta} -6 \lambda_{\Phi} v \sin^3{\theta} + \mu_3 \cos^3{\theta} + 3\mu_{S\Phi} \cos{\theta}\sin^2{\theta} \, , \\
\mu^\prime_{Hh} &=& \lambda_{S\Phi} v \cos^3{\theta} -2 \lambda_{S\Phi} v \cos{\theta}\sin^2{\theta}  +6\lambda_\Phi v \cos{\theta} \sin^2{\theta} \nonumber \\ && + \mu_3 \cos^2{\theta} \sin{\theta} - 2 \mu_{S\Phi} \cos^2{\theta} \sin{\theta} +\mu_{S\Phi} \sin^3{\theta}\, , \\
\mu_{Hh} &=& 2 \lambda_{S\Phi} v \cos^2{\theta} \sin{\theta} - \lambda_{S\Phi} v \sin^3{\theta} -6 \lambda_{\Phi} v \cos^2{\theta} \sin{\theta} \nonumber \\ && + \mu_3 \cos{\theta} \sin^2{\theta} +\mu_{S\Phi} \cos^3{\theta} -2 \mu_{S\Phi} \cos{\theta} \sin^2{\theta}  \, ,  \\
\mu_h &=& 3\lambda_{S\Phi} v \cos{\theta}\sin^2{\theta} + 6\lambda_\Phi v \cos^3{\theta} + \mu_3 \sin^3{\theta} + 3\mu_{S\Phi} \cos^2{\theta}\sin{\theta} \, , \\
\lambda_{\eta H} &=& \lambda_{\eta S}\cos^2{\theta}+\lambda_{\eta \Phi} \sin^2{\theta} \, ,\\
\lambda_{\eta h} &=& \lambda_{\eta S}\sin^2{\theta}+\lambda_{\eta \Phi} \cos^2{\theta} \, ,\\
\lambda_{\eta Hh} &=& \lambda_{\eta S}\cos{\theta}\sin{\theta}-\lambda_{\eta \Phi} \cos{\theta}\sin{\theta}\, ,\\
\lambda_0 &=& \lambda_{S} \cos^4{\theta} + 6 \lambda_{S\Phi}  \cos^2{\theta} \sin^2{\theta} + 6 \lambda_{\Phi} \sin^4{\theta} \, , \\
\lambda_1 &=& \lambda_S \cos^3{\theta} -3 \lambda_{S\Phi} \cos^3{\theta} \sin{\theta} + 3 \lambda_{S\Phi} \cos{\theta} \sin^3{\theta} -6 \lambda_{\Phi}\cos{\theta} \sin^3{\theta} \, , \\
\lambda_2 &=& \lambda_S \cos^2{\theta} \sin^2{\theta} +\lambda_{S\Phi} \cos^4{\theta} -4\lambda_{S\Phi} \cos^2{\theta} \sin^2{\theta} +\lambda_{S\Phi}\sin^4{\theta} +6 \lambda_{\Phi} \cos^2{\theta}\sin^2{\theta} \, , \\
\lambda_3 &=& \lambda_S \cos{\theta} \sin^3{\theta} +3 \lambda_{S\Phi} \cos^3{\theta} \sin{\theta} -3\lambda_{S\Phi} \cos{\theta} \sin^3{\theta} - 6\lambda_{\Phi} \cos^3{\theta} \sin{\theta} \, , \\
\lambda_4 &=& \lambda_{S} \sin^4{\theta} + 6 \lambda_{S\Phi} \cos^2{\theta} \sin^2{\theta} +6 \lambda_\Phi \cos^4{\theta} \, .
\end{eqnarray}

\bibliographystyle{apsrev4-1}
\bibliography{refs}

\begin{thebibliography}{93}%
\makeatletter
\providecommand \@ifxundefined [1]{%
 \@ifx{#1\undefined}
}%
\providecommand \@ifnum [1]{%
 \ifnum #1\expandafter \@firstoftwo
 \else \expandafter \@secondoftwo
 \fi
}%
\providecommand \@ifx [1]{%
 \ifx #1\expandafter \@firstoftwo
 \else \expandafter \@secondoftwo
 \fi
}%
\providecommand \natexlab [1]{#1}%
\providecommand \enquote  [1]{``#1''}%
\providecommand \bibnamefont  [1]{#1}%
\providecommand \bibfnamefont [1]{#1}%
\providecommand \citenamefont [1]{#1}%
\providecommand \href@noop [0]{\@secondoftwo}%
\providecommand \href [0]{\begingroup \@sanitize@url \@href}%
\providecommand \@href[1]{\@@startlink{#1}\@@href}%
\providecommand \@@href[1]{\endgroup#1\@@endlink}%
\providecommand \@sanitize@url [0]{\catcode `\\12\catcode `\$12\catcode
  `\&12\catcode `\#12\catcode `\^12\catcode `\_12\catcode `\%12\relax}%
\providecommand \@@startlink[1]{}%
\providecommand \@@endlink[0]{}%
\providecommand \url  [0]{\begingroup\@sanitize@url \@url }%
\providecommand \@url [1]{\endgroup\@href {#1}{\urlprefix }}%
\providecommand \urlprefix  [0]{URL }%
\providecommand \Eprint [0]{\href }%
\providecommand \doibase [0]{http://dx.doi.org/}%
\providecommand \selectlanguage [0]{\@gobble}%
\providecommand \bibinfo  [0]{\@secondoftwo}%
\providecommand \bibfield  [0]{\@secondoftwo}%
\providecommand \translation [1]{[#1]}%
\providecommand \BibitemOpen [0]{}%
\providecommand \bibitemStop [0]{}%
\providecommand \bibitemNoStop [0]{.\EOS\space}%
\providecommand \EOS [0]{\spacefactor3000\relax}%
\providecommand \BibitemShut  [1]{\csname bibitem#1\endcsname}%
\let\auto@bib@innerbib\@empty
\bibitem [{\citenamefont {Aghanim}\ \emph {et~al.}(2020)\citenamefont {Aghanim}
  \emph {et~al.}}]{Planck:2018vyg}%
  \BibitemOpen
  \bibfield  {author} {\bibinfo {author} {\bibfnamefont {N.}~\bibnamefont
  {Aghanim}} \emph {et~al.} (\bibinfo {collaboration} {Planck}),\ }\href
  {\doibase 10.1051/0004-6361/201833910} {\bibfield  {journal} {\bibinfo
  {journal} {Astron. Astrophys.}\ }\textbf {\bibinfo {volume} {641}},\ \bibinfo
  {pages} {A6} (\bibinfo {year} {2020})},\ \Eprint
  {http://arxiv.org/abs/1807.06209} {arXiv:1807.06209 [astro-ph.CO]}
  \BibitemShut {NoStop}%
\bibitem [{\citenamefont {Aprile}\ \emph {et~al.}(2018)\citenamefont {Aprile}
  \emph {et~al.}}]{XENON:2018voc}%
  \BibitemOpen
  \bibfield  {author} {\bibinfo {author} {\bibfnamefont {E.}~\bibnamefont
  {Aprile}} \emph {et~al.} (\bibinfo {collaboration} {XENON}),\ }\href
  {\doibase 10.1103/PhysRevLett.121.111302} {\bibfield  {journal} {\bibinfo
  {journal} {Phys. Rev. Lett.}\ }\textbf {\bibinfo {volume} {121}},\ \bibinfo
  {pages} {111302} (\bibinfo {year} {2018})},\ \Eprint
  {http://arxiv.org/abs/1805.12562} {arXiv:1805.12562 [astro-ph.CO]}
  \BibitemShut {NoStop}%
\bibitem [{\citenamefont {Akerib}\ \emph {et~al.}(2017)\citenamefont {Akerib}
  \emph {et~al.}}]{LUX:2016ggv}%
  \BibitemOpen
  \bibfield  {author} {\bibinfo {author} {\bibfnamefont {D.~S.}\ \bibnamefont
  {Akerib}} \emph {et~al.} (\bibinfo {collaboration} {LUX}),\ }\href {\doibase
  10.1103/PhysRevLett.118.021303} {\bibfield  {journal} {\bibinfo  {journal}
  {Phys. Rev. Lett.}\ }\textbf {\bibinfo {volume} {118}},\ \bibinfo {pages}
  {021303} (\bibinfo {year} {2017})},\ \Eprint
  {http://arxiv.org/abs/1608.07648} {arXiv:1608.07648 [astro-ph.CO]}
  \BibitemShut {NoStop}%
\bibitem [{\citenamefont {Lee}\ and\ \citenamefont
  {Weinberg}(1977)}]{Lee:1977ua}%
  \BibitemOpen
  \bibfield  {author} {\bibinfo {author} {\bibfnamefont {B.~W.}\ \bibnamefont
  {Lee}}\ and\ \bibinfo {author} {\bibfnamefont {S.}~\bibnamefont {Weinberg}},\
  }\href {\doibase 10.1103/PhysRevLett.39.165} {\bibfield  {journal} {\bibinfo
  {journal} {Phys. Rev. Lett.}\ }\textbf {\bibinfo {volume} {39}},\ \bibinfo
  {pages} {165} (\bibinfo {year} {1977})}\BibitemShut {NoStop}%
\bibitem [{\citenamefont {Silveira}\ and\ \citenamefont
  {Zee}(1985)}]{Silveira:1985rk}%
  \BibitemOpen
  \bibfield  {author} {\bibinfo {author} {\bibfnamefont {V.}~\bibnamefont
  {Silveira}}\ and\ \bibinfo {author} {\bibfnamefont {A.}~\bibnamefont {Zee}},\
  }\href {\doibase 10.1016/0370-2693(85)90624-0} {\bibfield  {journal}
  {\bibinfo  {journal} {Phys. Lett. B}\ }\textbf {\bibinfo {volume} {161}},\
  \bibinfo {pages} {136} (\bibinfo {year} {1985})}\BibitemShut {NoStop}%
\bibitem [{\citenamefont {McDonald}(1994)}]{McDonald:1993ex}%
  \BibitemOpen
  \bibfield  {author} {\bibinfo {author} {\bibfnamefont {J.}~\bibnamefont
  {McDonald}},\ }\href {\doibase 10.1103/PhysRevD.50.3637} {\bibfield
  {journal} {\bibinfo  {journal} {Phys. Rev. D}\ }\textbf {\bibinfo {volume}
  {50}},\ \bibinfo {pages} {3637} (\bibinfo {year} {1994})},\ \Eprint
  {http://arxiv.org/abs/hep-ph/0702143} {arXiv:hep-ph/0702143} \BibitemShut
  {NoStop}%
\bibitem [{\citenamefont {Burgess}\ \emph {et~al.}(2001)\citenamefont
  {Burgess}, \citenamefont {Pospelov},\ and\ \citenamefont {ter
  Veldhuis}}]{Burgess:2000yq}%
  \BibitemOpen
  \bibfield  {author} {\bibinfo {author} {\bibfnamefont {C.~P.}\ \bibnamefont
  {Burgess}}, \bibinfo {author} {\bibfnamefont {M.}~\bibnamefont {Pospelov}}, \
  and\ \bibinfo {author} {\bibfnamefont {T.}~\bibnamefont {ter Veldhuis}},\
  }\href {\doibase 10.1016/S0550-3213(01)00513-2} {\bibfield  {journal}
  {\bibinfo  {journal} {Nucl. Phys. B}\ }\textbf {\bibinfo {volume} {619}},\
  \bibinfo {pages} {709} (\bibinfo {year} {2001})},\ \Eprint
  {http://arxiv.org/abs/hep-ph/0011335} {arXiv:hep-ph/0011335} \BibitemShut
  {NoStop}%
\bibitem [{\citenamefont {Barger}\ \emph {et~al.}(2008)\citenamefont {Barger},
  \citenamefont {Langacker}, \citenamefont {McCaskey}, \citenamefont
  {Ramsey-Musolf},\ and\ \citenamefont {Shaughnessy}}]{Barger:2007im}%
  \BibitemOpen
  \bibfield  {author} {\bibinfo {author} {\bibfnamefont {V.}~\bibnamefont
  {Barger}}, \bibinfo {author} {\bibfnamefont {P.}~\bibnamefont {Langacker}},
  \bibinfo {author} {\bibfnamefont {M.}~\bibnamefont {McCaskey}}, \bibinfo
  {author} {\bibfnamefont {M.~J.}\ \bibnamefont {Ramsey-Musolf}}, \ and\
  \bibinfo {author} {\bibfnamefont {G.}~\bibnamefont {Shaughnessy}},\ }\href
  {\doibase 10.1103/PhysRevD.77.035005} {\bibfield  {journal} {\bibinfo
  {journal} {Phys. Rev. D}\ }\textbf {\bibinfo {volume} {77}},\ \bibinfo
  {pages} {035005} (\bibinfo {year} {2008})},\ \Eprint
  {http://arxiv.org/abs/0706.4311} {arXiv:0706.4311 [hep-ph]} \BibitemShut
  {NoStop}%
\bibitem [{\citenamefont {Lerner}\ and\ \citenamefont
  {McDonald}(2009)}]{Lerner:2009xg}%
  \BibitemOpen
  \bibfield  {author} {\bibinfo {author} {\bibfnamefont {R.~N.}\ \bibnamefont
  {Lerner}}\ and\ \bibinfo {author} {\bibfnamefont {J.}~\bibnamefont
  {McDonald}},\ }\href {\doibase 10.1103/PhysRevD.80.123507} {\bibfield
  {journal} {\bibinfo  {journal} {Phys. Rev. D}\ }\textbf {\bibinfo {volume}
  {80}},\ \bibinfo {pages} {123507} (\bibinfo {year} {2009})},\ \Eprint
  {http://arxiv.org/abs/0909.0520} {arXiv:0909.0520 [hep-ph]} \BibitemShut
  {NoStop}%
\bibitem [{\citenamefont {Athron}\ \emph {et~al.}(2017)\citenamefont {Athron}
  \emph {et~al.}}]{GAMBIT:2017gge}%
  \BibitemOpen
  \bibfield  {author} {\bibinfo {author} {\bibfnamefont {P.}~\bibnamefont
  {Athron}} \emph {et~al.} (\bibinfo {collaboration} {GAMBIT}),\ }\href
  {\doibase 10.1140/epjc/s10052-017-5113-1} {\bibfield  {journal} {\bibinfo
  {journal} {Eur. Phys. J. C}\ }\textbf {\bibinfo {volume} {77}},\ \bibinfo
  {pages} {568} (\bibinfo {year} {2017})},\ \Eprint
  {http://arxiv.org/abs/1705.07931} {arXiv:1705.07931 [hep-ph]} \BibitemShut
  {NoStop}%
\bibitem [{\citenamefont {Kanemura}\ \emph {et~al.}(2010)\citenamefont
  {Kanemura}, \citenamefont {Matsumoto}, \citenamefont {Nabeshima},\ and\
  \citenamefont {Okada}}]{Kanemura:2010sh}%
  \BibitemOpen
  \bibfield  {author} {\bibinfo {author} {\bibfnamefont {S.}~\bibnamefont
  {Kanemura}}, \bibinfo {author} {\bibfnamefont {S.}~\bibnamefont {Matsumoto}},
  \bibinfo {author} {\bibfnamefont {T.}~\bibnamefont {Nabeshima}}, \ and\
  \bibinfo {author} {\bibfnamefont {N.}~\bibnamefont {Okada}},\ }\href
  {\doibase 10.1103/PhysRevD.82.055026} {\bibfield  {journal} {\bibinfo
  {journal} {Phys. Rev. D}\ }\textbf {\bibinfo {volume} {82}},\ \bibinfo
  {pages} {055026} (\bibinfo {year} {2010})},\ \Eprint
  {http://arxiv.org/abs/1005.5651} {arXiv:1005.5651 [hep-ph]} \BibitemShut
  {NoStop}%
\bibitem [{\citenamefont {Cline}\ \emph {et~al.}(2013)\citenamefont {Cline},
  \citenamefont {Kainulainen}, \citenamefont {Scott},\ and\ \citenamefont
  {Weniger}}]{Cline:2013gha}%
  \BibitemOpen
  \bibfield  {author} {\bibinfo {author} {\bibfnamefont {J.~M.}\ \bibnamefont
  {Cline}}, \bibinfo {author} {\bibfnamefont {K.}~\bibnamefont {Kainulainen}},
  \bibinfo {author} {\bibfnamefont {P.}~\bibnamefont {Scott}}, \ and\ \bibinfo
  {author} {\bibfnamefont {C.}~\bibnamefont {Weniger}},\ }\href {\doibase
  10.1103/PhysRevD.88.055025} {\bibfield  {journal} {\bibinfo  {journal} {Phys.
  Rev. D}\ }\textbf {\bibinfo {volume} {88}},\ \bibinfo {pages} {055025}
  (\bibinfo {year} {2013})},\ \bibinfo {note} {[Erratum: Phys.Rev.D 92, 039906
  (2015)]},\ \Eprint {http://arxiv.org/abs/1306.4710} {arXiv:1306.4710
  [hep-ph]} \BibitemShut {NoStop}%
\bibitem [{\citenamefont {Dolan}\ \emph {et~al.}(2015)\citenamefont {Dolan},
  \citenamefont {Kahlhoefer}, \citenamefont {McCabe},\ and\ \citenamefont
  {Schmidt-Hoberg}}]{Dolan:2014ska}%
  \BibitemOpen
  \bibfield  {author} {\bibinfo {author} {\bibfnamefont {M.~J.}\ \bibnamefont
  {Dolan}}, \bibinfo {author} {\bibfnamefont {F.}~\bibnamefont {Kahlhoefer}},
  \bibinfo {author} {\bibfnamefont {C.}~\bibnamefont {McCabe}}, \ and\ \bibinfo
  {author} {\bibfnamefont {K.}~\bibnamefont {Schmidt-Hoberg}},\ }\href
  {\doibase 10.1007/JHEP03(2015)171} {\bibfield  {journal} {\bibinfo  {journal}
  {JHEP}\ }\textbf {\bibinfo {volume} {03}},\ \bibinfo {pages} {171} (\bibinfo
  {year} {2015})},\ \bibinfo {note} {[Erratum: JHEP 07, 103 (2015)]},\ \Eprint
  {http://arxiv.org/abs/1412.5174} {arXiv:1412.5174 [hep-ph]} \BibitemShut
  {NoStop}%
\bibitem [{\citenamefont {Krnjaic}(2016)}]{Krnjaic:2015mbs}%
  \BibitemOpen
  \bibfield  {author} {\bibinfo {author} {\bibfnamefont {G.}~\bibnamefont
  {Krnjaic}},\ }\href {\doibase 10.1103/PhysRevD.94.073009} {\bibfield
  {journal} {\bibinfo  {journal} {Phys. Rev. D}\ }\textbf {\bibinfo {volume}
  {94}},\ \bibinfo {pages} {073009} (\bibinfo {year} {2016})},\ \Eprint
  {http://arxiv.org/abs/1512.04119} {arXiv:1512.04119 [hep-ph]} \BibitemShut
  {NoStop}%
\bibitem [{\citenamefont {Matsumoto}\ \emph {et~al.}(2019)\citenamefont
  {Matsumoto}, \citenamefont {Tsai},\ and\ \citenamefont
  {Tseng}}]{Matsumoto:2018acr}%
  \BibitemOpen
  \bibfield  {author} {\bibinfo {author} {\bibfnamefont {S.}~\bibnamefont
  {Matsumoto}}, \bibinfo {author} {\bibfnamefont {Y.-L.~S.}\ \bibnamefont
  {Tsai}}, \ and\ \bibinfo {author} {\bibfnamefont {P.-Y.}\ \bibnamefont
  {Tseng}},\ }\href {\doibase 10.1007/JHEP07(2019)050} {\bibfield  {journal}
  {\bibinfo  {journal} {JHEP}\ }\textbf {\bibinfo {volume} {07}},\ \bibinfo
  {pages} {050} (\bibinfo {year} {2019})},\ \Eprint
  {http://arxiv.org/abs/1811.03292} {arXiv:1811.03292 [hep-ph]} \BibitemShut
  {NoStop}%
\bibitem [{\citenamefont {Bondarenko}\ \emph {et~al.}(2020)\citenamefont
  {Bondarenko}, \citenamefont {Boyarsky}, \citenamefont {Bringmann},
  \citenamefont {Hufnagel}, \citenamefont {Schmidt-Hoberg},\ and\ \citenamefont
  {Sokolenko}}]{Bondarenko:2019vrb}%
  \BibitemOpen
  \bibfield  {author} {\bibinfo {author} {\bibfnamefont {K.}~\bibnamefont
  {Bondarenko}}, \bibinfo {author} {\bibfnamefont {A.}~\bibnamefont
  {Boyarsky}}, \bibinfo {author} {\bibfnamefont {T.}~\bibnamefont {Bringmann}},
  \bibinfo {author} {\bibfnamefont {M.}~\bibnamefont {Hufnagel}}, \bibinfo
  {author} {\bibfnamefont {K.}~\bibnamefont {Schmidt-Hoberg}}, \ and\ \bibinfo
  {author} {\bibfnamefont {A.}~\bibnamefont {Sokolenko}},\ }\href {\doibase
  10.1007/JHEP03(2020)118} {\bibfield  {journal} {\bibinfo  {journal} {JHEP}\
  }\textbf {\bibinfo {volume} {03}},\ \bibinfo {pages} {118} (\bibinfo {year}
  {2020})},\ \Eprint {http://arxiv.org/abs/1909.08632} {arXiv:1909.08632
  [hep-ph]} \BibitemShut {NoStop}%
\bibitem [{\citenamefont {Coito}\ \emph {et~al.}(2021)\citenamefont {Coito},
  \citenamefont {Faubel}, \citenamefont {Herrero-Garcia},\ and\ \citenamefont
  {Santamaria}}]{Coito:2021fgo}%
  \BibitemOpen
  \bibfield  {author} {\bibinfo {author} {\bibfnamefont {L.}~\bibnamefont
  {Coito}}, \bibinfo {author} {\bibfnamefont {C.}~\bibnamefont {Faubel}},
  \bibinfo {author} {\bibfnamefont {J.}~\bibnamefont {Herrero-Garcia}}, \ and\
  \bibinfo {author} {\bibfnamefont {A.}~\bibnamefont {Santamaria}},\
  }\href@noop {} {\  (\bibinfo {year} {2021})},\ \Eprint
  {http://arxiv.org/abs/2106.05289} {arXiv:2106.05289 [hep-ph]} \BibitemShut
  {NoStop}%
\bibitem [{\citenamefont {Abada}\ \emph {et~al.}(2011)\citenamefont {Abada},
  \citenamefont {Ghaffor},\ and\ \citenamefont {Nasri}}]{Abada:2011qb}%
  \BibitemOpen
  \bibfield  {author} {\bibinfo {author} {\bibfnamefont {A.}~\bibnamefont
  {Abada}}, \bibinfo {author} {\bibfnamefont {D.}~\bibnamefont {Ghaffor}}, \
  and\ \bibinfo {author} {\bibfnamefont {S.}~\bibnamefont {Nasri}},\ }\href
  {\doibase 10.1103/PhysRevD.83.095021} {\bibfield  {journal} {\bibinfo
  {journal} {Phys. Rev. D}\ }\textbf {\bibinfo {volume} {83}},\ \bibinfo
  {pages} {095021} (\bibinfo {year} {2011})},\ \Eprint
  {http://arxiv.org/abs/1101.0365} {arXiv:1101.0365 [hep-ph]} \BibitemShut
  {NoStop}%
\bibitem [{\citenamefont {Abada}\ and\ \citenamefont
  {Nasri}(2012)}]{Abada:2012hf}%
  \BibitemOpen
  \bibfield  {author} {\bibinfo {author} {\bibfnamefont {A.}~\bibnamefont
  {Abada}}\ and\ \bibinfo {author} {\bibfnamefont {S.}~\bibnamefont {Nasri}},\
  }\href {\doibase 10.1103/PhysRevD.85.075009} {\bibfield  {journal} {\bibinfo
  {journal} {Phys. Rev. D}\ }\textbf {\bibinfo {volume} {85}},\ \bibinfo
  {pages} {075009} (\bibinfo {year} {2012})},\ \Eprint
  {http://arxiv.org/abs/1201.1413} {arXiv:1201.1413 [hep-ph]} \BibitemShut
  {NoStop}%
\bibitem [{\citenamefont {Ahriche}\ \emph {et~al.}(2014)\citenamefont
  {Ahriche}, \citenamefont {Arhrib},\ and\ \citenamefont
  {Nasri}}]{Ahriche:2013vqa}%
  \BibitemOpen
  \bibfield  {author} {\bibinfo {author} {\bibfnamefont {A.}~\bibnamefont
  {Ahriche}}, \bibinfo {author} {\bibfnamefont {A.}~\bibnamefont {Arhrib}}, \
  and\ \bibinfo {author} {\bibfnamefont {S.}~\bibnamefont {Nasri}},\ }\href
  {\doibase 10.1007/JHEP02(2014)042} {\bibfield  {journal} {\bibinfo  {journal}
  {JHEP}\ }\textbf {\bibinfo {volume} {02}},\ \bibinfo {pages} {042} (\bibinfo
  {year} {2014})},\ \Eprint {http://arxiv.org/abs/1309.5615} {arXiv:1309.5615
  [hep-ph]} \BibitemShut {NoStop}%
\bibitem [{\citenamefont {Arhrib}\ and\ \citenamefont
  {Maniatis}(2019)}]{Arhrib:2018eex}%
  \BibitemOpen
  \bibfield  {author} {\bibinfo {author} {\bibfnamefont {A.}~\bibnamefont
  {Arhrib}}\ and\ \bibinfo {author} {\bibfnamefont {M.}~\bibnamefont
  {Maniatis}},\ }\href {\doibase 10.1016/j.physletb.2019.07.023} {\bibfield
  {journal} {\bibinfo  {journal} {Phys. Lett. B}\ }\textbf {\bibinfo {volume}
  {796}},\ \bibinfo {pages} {15} (\bibinfo {year} {2019})},\ \Eprint
  {http://arxiv.org/abs/1807.03554} {arXiv:1807.03554 [hep-ph]} \BibitemShut
  {NoStop}%
\bibitem [{\citenamefont {Maniatis}(2021)}]{Maniatis:2020ois}%
  \BibitemOpen
  \bibfield  {author} {\bibinfo {author} {\bibfnamefont {M.}~\bibnamefont
  {Maniatis}},\ }\href {\doibase 10.1103/PhysRevD.103.015010} {\bibfield
  {journal} {\bibinfo  {journal} {Phys. Rev. D}\ }\textbf {\bibinfo {volume}
  {103}},\ \bibinfo {pages} {015010} (\bibinfo {year} {2021})},\ \Eprint
  {http://arxiv.org/abs/2005.13443} {arXiv:2005.13443 [hep-ph]} \BibitemShut
  {NoStop}%
\bibitem [{\citenamefont {Basak}\ \emph {et~al.}(2021)\citenamefont {Basak},
  \citenamefont {Coleppa},\ and\ \citenamefont {Loho}}]{Basak:2021tnj}%
  \BibitemOpen
  \bibfield  {author} {\bibinfo {author} {\bibfnamefont {T.}~\bibnamefont
  {Basak}}, \bibinfo {author} {\bibfnamefont {B.}~\bibnamefont {Coleppa}}, \
  and\ \bibinfo {author} {\bibfnamefont {K.}~\bibnamefont {Loho}},\ }\href
  {\doibase 10.1007/JHEP06(2021)104} {\bibfield  {journal} {\bibinfo  {journal}
  {JHEP}\ }\textbf {\bibinfo {volume} {06}},\ \bibinfo {pages} {104} (\bibinfo
  {year} {2021})},\ \Eprint {http://arxiv.org/abs/2105.09044} {arXiv:2105.09044
  [hep-ph]} \BibitemShut {NoStop}%
\bibitem [{\citenamefont {D'Agnolo}\ and\ \citenamefont
  {Ruderman}(2015)}]{DAgnolo:2015ujb}%
  \BibitemOpen
  \bibfield  {author} {\bibinfo {author} {\bibfnamefont {R.~T.}\ \bibnamefont
  {D'Agnolo}}\ and\ \bibinfo {author} {\bibfnamefont {J.~T.}\ \bibnamefont
  {Ruderman}},\ }\href {\doibase 10.1103/PhysRevLett.115.061301} {\bibfield
  {journal} {\bibinfo  {journal} {Phys. Rev. Lett.}\ }\textbf {\bibinfo
  {volume} {115}},\ \bibinfo {pages} {061301} (\bibinfo {year} {2015})},\
  \Eprint {http://arxiv.org/abs/1505.07107} {arXiv:1505.07107 [hep-ph]}
  \BibitemShut {NoStop}%
\bibitem [{\citenamefont {Okada}\ and\ \citenamefont
  {Seto}(2020)}]{Okada:2019sbb}%
  \BibitemOpen
  \bibfield  {author} {\bibinfo {author} {\bibfnamefont {N.}~\bibnamefont
  {Okada}}\ and\ \bibinfo {author} {\bibfnamefont {O.}~\bibnamefont {Seto}},\
  }\href {\doibase 10.1103/PhysRevD.101.023522} {\bibfield  {journal} {\bibinfo
   {journal} {Phys. Rev. D}\ }\textbf {\bibinfo {volume} {101}},\ \bibinfo
  {pages} {023522} (\bibinfo {year} {2020})},\ \Eprint
  {http://arxiv.org/abs/1908.09277} {arXiv:1908.09277 [hep-ph]} \BibitemShut
  {NoStop}%
\bibitem [{\citenamefont {Griest}\ and\ \citenamefont
  {Seckel}(1991)}]{Griest:1990kh}%
  \BibitemOpen
  \bibfield  {author} {\bibinfo {author} {\bibfnamefont {K.}~\bibnamefont
  {Griest}}\ and\ \bibinfo {author} {\bibfnamefont {D.}~\bibnamefont
  {Seckel}},\ }\href {\doibase 10.1103/PhysRevD.43.3191} {\bibfield  {journal}
  {\bibinfo  {journal} {Phys. Rev. D}\ }\textbf {\bibinfo {volume} {43}},\
  \bibinfo {pages} {3191} (\bibinfo {year} {1991})}\BibitemShut {NoStop}%
\bibitem [{\citenamefont {Gunion}\ \emph {et~al.}(2000)\citenamefont {Gunion},
  \citenamefont {Haber}, \citenamefont {Kane},\ and\ \citenamefont
  {Dawson}}]{Gunion:1989we}%
  \BibitemOpen
  \bibfield  {author} {\bibinfo {author} {\bibfnamefont {J.~F.}\ \bibnamefont
  {Gunion}}, \bibinfo {author} {\bibfnamefont {H.~E.}\ \bibnamefont {Haber}},
  \bibinfo {author} {\bibfnamefont {G.~L.}\ \bibnamefont {Kane}}, \ and\
  \bibinfo {author} {\bibfnamefont {S.}~\bibnamefont {Dawson}},\ }\href@noop {}
  {\emph {\bibinfo {title} {{The Higgs Hunter's Guide}}}},\ Vol.~\bibinfo
  {volume} {80}\ (\bibinfo {year} {2000})\BibitemShut {NoStop}%
\bibitem [{\citenamefont {Raby}\ and\ \citenamefont
  {West}(1988)}]{Raby:1988qf}%
  \BibitemOpen
  \bibfield  {author} {\bibinfo {author} {\bibfnamefont {S.}~\bibnamefont
  {Raby}}\ and\ \bibinfo {author} {\bibfnamefont {G.~B.}\ \bibnamefont
  {West}},\ }\href {\doibase 10.1103/PhysRevD.38.3488} {\bibfield  {journal}
  {\bibinfo  {journal} {Phys. Rev. D}\ }\textbf {\bibinfo {volume} {38}},\
  \bibinfo {pages} {3488} (\bibinfo {year} {1988})}\BibitemShut {NoStop}%
\bibitem [{\citenamefont {Monin}\ \emph {et~al.}(2019)\citenamefont {Monin},
  \citenamefont {Boyarsky},\ and\ \citenamefont {Ruchayskiy}}]{Monin:2018lee}%
  \BibitemOpen
  \bibfield  {author} {\bibinfo {author} {\bibfnamefont {A.}~\bibnamefont
  {Monin}}, \bibinfo {author} {\bibfnamefont {A.}~\bibnamefont {Boyarsky}}, \
  and\ \bibinfo {author} {\bibfnamefont {O.}~\bibnamefont {Ruchayskiy}},\
  }\href {\doibase 10.1103/PhysRevD.99.015019} {\bibfield  {journal} {\bibinfo
  {journal} {Phys. Rev. D}\ }\textbf {\bibinfo {volume} {99}},\ \bibinfo
  {pages} {015019} (\bibinfo {year} {2019})},\ \Eprint
  {http://arxiv.org/abs/1806.07759} {arXiv:1806.07759 [hep-ph]} \BibitemShut
  {NoStop}%
\bibitem [{\citenamefont {Winkler}(2019)}]{Winkler:2018qyg}%
  \BibitemOpen
  \bibfield  {author} {\bibinfo {author} {\bibfnamefont {M.~W.}\ \bibnamefont
  {Winkler}},\ }\href {\doibase 10.1103/PhysRevD.99.015018} {\bibfield
  {journal} {\bibinfo  {journal} {Phys. Rev. D}\ }\textbf {\bibinfo {volume}
  {99}},\ \bibinfo {pages} {015018} (\bibinfo {year} {2019})},\ \Eprint
  {http://arxiv.org/abs/1809.01876} {arXiv:1809.01876 [hep-ph]} \BibitemShut
  {NoStop}%
\bibitem [{\citenamefont {Kannike}(2012)}]{Kannike:2012pe}%
  \BibitemOpen
  \bibfield  {author} {\bibinfo {author} {\bibfnamefont {K.}~\bibnamefont
  {Kannike}},\ }\href {\doibase 10.1140/epjc/s10052-012-2093-z} {\bibfield
  {journal} {\bibinfo  {journal} {Eur. Phys. J. C}\ }\textbf {\bibinfo {volume}
  {72}},\ \bibinfo {pages} {2093} (\bibinfo {year} {2012})},\ \Eprint
  {http://arxiv.org/abs/1205.3781} {arXiv:1205.3781 [hep-ph]} \BibitemShut
  {NoStop}%
\bibitem [{\citenamefont {Lee}\ \emph {et~al.}(1977)\citenamefont {Lee},
  \citenamefont {Quigg},\ and\ \citenamefont {Thacker}}]{Lee:1977eg}%
  \BibitemOpen
  \bibfield  {author} {\bibinfo {author} {\bibfnamefont {B.~W.}\ \bibnamefont
  {Lee}}, \bibinfo {author} {\bibfnamefont {C.}~\bibnamefont {Quigg}}, \ and\
  \bibinfo {author} {\bibfnamefont {H.~B.}\ \bibnamefont {Thacker}},\ }\href
  {\doibase 10.1103/PhysRevD.16.1519} {\bibfield  {journal} {\bibinfo
  {journal} {Phys. Rev. D}\ }\textbf {\bibinfo {volume} {16}},\ \bibinfo
  {pages} {1519} (\bibinfo {year} {1977})}\BibitemShut {NoStop}%
\bibitem [{\citenamefont {Marciano}\ \emph {et~al.}(1989)\citenamefont
  {Marciano}, \citenamefont {Valencia},\ and\ \citenamefont
  {Willenbrock}}]{Marciano:1989ns}%
  \BibitemOpen
  \bibfield  {author} {\bibinfo {author} {\bibfnamefont {W.~J.}\ \bibnamefont
  {Marciano}}, \bibinfo {author} {\bibfnamefont {G.}~\bibnamefont {Valencia}},
  \ and\ \bibinfo {author} {\bibfnamefont {S.}~\bibnamefont {Willenbrock}},\
  }\href {\doibase 10.1103/PhysRevD.40.1725} {\bibfield  {journal} {\bibinfo
  {journal} {Phys. Rev. D}\ }\textbf {\bibinfo {volume} {40}},\ \bibinfo
  {pages} {1725} (\bibinfo {year} {1989})}\BibitemShut {NoStop}%
\bibitem [{\citenamefont {Cornwall}\ \emph {et~al.}(1974)\citenamefont
  {Cornwall}, \citenamefont {Levin},\ and\ \citenamefont
  {Tiktopoulos}}]{Cornwall:1974km}%
  \BibitemOpen
  \bibfield  {author} {\bibinfo {author} {\bibfnamefont {J.~M.}\ \bibnamefont
  {Cornwall}}, \bibinfo {author} {\bibfnamefont {D.~N.}\ \bibnamefont {Levin}},
  \ and\ \bibinfo {author} {\bibfnamefont {G.}~\bibnamefont {Tiktopoulos}},\
  }\href {\doibase 10.1103/PhysRevD.10.1145} {\bibfield  {journal} {\bibinfo
  {journal} {Phys. Rev. D}\ }\textbf {\bibinfo {volume} {10}},\ \bibinfo
  {pages} {1145} (\bibinfo {year} {1974})},\ \bibinfo {note} {[Erratum:
  Phys.Rev.D 11, 972 (1975)]}\BibitemShut {NoStop}%
\bibitem [{\citenamefont {Gondolo}\ and\ \citenamefont
  {Gelmini}(1991)}]{Gondolo:1990dk}%
  \BibitemOpen
  \bibfield  {author} {\bibinfo {author} {\bibfnamefont {P.}~\bibnamefont
  {Gondolo}}\ and\ \bibinfo {author} {\bibfnamefont {G.}~\bibnamefont
  {Gelmini}},\ }\href {\doibase 10.1016/0550-3213(91)90438-4} {\bibfield
  {journal} {\bibinfo  {journal} {Nucl. Phys. B}\ }\textbf {\bibinfo {volume}
  {360}},\ \bibinfo {pages} {145} (\bibinfo {year} {1991})}\BibitemShut
  {NoStop}%
\bibitem [{\citenamefont {Acciarri}\ \emph {et~al.}(1996)\citenamefont
  {Acciarri} \emph {et~al.}}]{L3:1996ome}%
  \BibitemOpen
  \bibfield  {author} {\bibinfo {author} {\bibfnamefont {M.}~\bibnamefont
  {Acciarri}} \emph {et~al.} (\bibinfo {collaboration} {L3}),\ }\href {\doibase
  10.1016/0370-2693(96)00987-2} {\bibfield  {journal} {\bibinfo  {journal}
  {Phys. Lett. B}\ }\textbf {\bibinfo {volume} {385}},\ \bibinfo {pages} {454}
  (\bibinfo {year} {1996})}\BibitemShut {NoStop}%
\bibitem [{\citenamefont {Zyla}\ \emph {et~al.}(2020)\citenamefont {Zyla} \emph
  {et~al.}}]{ParticleDataGroup:2020ssz}%
  \BibitemOpen
  \bibfield  {author} {\bibinfo {author} {\bibfnamefont {P.~A.}\ \bibnamefont
  {Zyla}} \emph {et~al.} (\bibinfo {collaboration} {Particle Data Group}),\
  }\href {\doibase 10.1093/ptep/ptaa104} {\bibfield  {journal} {\bibinfo
  {journal} {PTEP}\ }\textbf {\bibinfo {volume} {2020}},\ \bibinfo {pages}
  {083C01} (\bibinfo {year} {2020})}\BibitemShut {NoStop}%
\bibitem [{\citenamefont {Aaboud}\ \emph {et~al.}(2019)\citenamefont {Aaboud}
  \emph {et~al.}}]{ATLAS:2018bnv}%
  \BibitemOpen
  \bibfield  {author} {\bibinfo {author} {\bibfnamefont {M.}~\bibnamefont
  {Aaboud}} \emph {et~al.} (\bibinfo {collaboration} {ATLAS}),\ }\href
  {\doibase 10.1016/j.physletb.2019.04.024} {\bibfield  {journal} {\bibinfo
  {journal} {Phys. Lett. B}\ }\textbf {\bibinfo {volume} {793}},\ \bibinfo
  {pages} {499} (\bibinfo {year} {2019})},\ \Eprint
  {http://arxiv.org/abs/1809.06682} {arXiv:1809.06682 [hep-ex]} \BibitemShut
  {NoStop}%
\bibitem [{\citenamefont {Wilczek}(1977)}]{Wilczek:1977zn}%
  \BibitemOpen
  \bibfield  {author} {\bibinfo {author} {\bibfnamefont {F.}~\bibnamefont
  {Wilczek}},\ }\href {\doibase 10.1103/PhysRevLett.39.1304} {\bibfield
  {journal} {\bibinfo  {journal} {Phys. Rev. Lett.}\ }\textbf {\bibinfo
  {volume} {39}},\ \bibinfo {pages} {1304} (\bibinfo {year}
  {1977})}\BibitemShut {NoStop}%
\bibitem [{\citenamefont {Adams}\ \emph {et~al.}(2005)\citenamefont {Adams}
  \emph {et~al.}}]{CLEO:2004tkr}%
  \BibitemOpen
  \bibfield  {author} {\bibinfo {author} {\bibfnamefont {G.~S.}\ \bibnamefont
  {Adams}} \emph {et~al.} (\bibinfo {collaboration} {CLEO}),\ }\href {\doibase
  10.1103/PhysRevLett.94.012001} {\bibfield  {journal} {\bibinfo  {journal}
  {Phys. Rev. Lett.}\ }\textbf {\bibinfo {volume} {94}},\ \bibinfo {pages}
  {012001} (\bibinfo {year} {2005})},\ \Eprint
  {http://arxiv.org/abs/hep-ex/0409027} {arXiv:hep-ex/0409027} \BibitemShut
  {NoStop}%
\bibitem [{\citenamefont {Seong}\ \emph {et~al.}(2019)\citenamefont {Seong}
  \emph {et~al.}}]{Belle:2018pzt}%
  \BibitemOpen
  \bibfield  {author} {\bibinfo {author} {\bibfnamefont {I.~S.}\ \bibnamefont
  {Seong}} \emph {et~al.} (\bibinfo {collaboration} {Belle}),\ }\href {\doibase
  10.1103/PhysRevLett.122.011801} {\bibfield  {journal} {\bibinfo  {journal}
  {Phys. Rev. Lett.}\ }\textbf {\bibinfo {volume} {122}},\ \bibinfo {pages}
  {011801} (\bibinfo {year} {2019})},\ \Eprint
  {http://arxiv.org/abs/1809.05222} {arXiv:1809.05222 [hep-ex]} \BibitemShut
  {NoStop}%
\bibitem [{\citenamefont {del Amo~Sanchez}\ \emph {et~al.}(2011)\citenamefont
  {del Amo~Sanchez} \emph {et~al.}}]{BaBar:2010eww}%
  \BibitemOpen
  \bibfield  {author} {\bibinfo {author} {\bibfnamefont {P.}~\bibnamefont {del
  Amo~Sanchez}} \emph {et~al.} (\bibinfo {collaboration} {BaBar}),\ }\href
  {\doibase 10.1103/PhysRevLett.107.021804} {\bibfield  {journal} {\bibinfo
  {journal} {Phys. Rev. Lett.}\ }\textbf {\bibinfo {volume} {107}},\ \bibinfo
  {pages} {021804} (\bibinfo {year} {2011})},\ \Eprint
  {http://arxiv.org/abs/1007.4646} {arXiv:1007.4646 [hep-ex]} \BibitemShut
  {NoStop}%
\bibitem [{\citenamefont {Aubert}\ \emph {et~al.}(2008)\citenamefont {Aubert}
  \emph {et~al.}}]{BaBar:2008aby}%
  \BibitemOpen
  \bibfield  {author} {\bibinfo {author} {\bibfnamefont {B.}~\bibnamefont
  {Aubert}} \emph {et~al.} (\bibinfo {collaboration} {BaBar}),\ }in\ \href@noop
  {} {\emph {\bibinfo {booktitle} {{34th International Conference on High
  Energy Physics}}}}\ (\bibinfo {year} {2008})\ \Eprint
  {http://arxiv.org/abs/0808.0017} {arXiv:0808.0017 [hep-ex]} \BibitemShut
  {NoStop}%
\bibitem [{\citenamefont {Love}\ \emph {et~al.}(2008)\citenamefont {Love} \emph
  {et~al.}}]{CLEO:2008jdl}%
  \BibitemOpen
  \bibfield  {author} {\bibinfo {author} {\bibfnamefont {W.}~\bibnamefont
  {Love}} \emph {et~al.} (\bibinfo {collaboration} {CLEO}),\ }\href {\doibase
  10.1103/PhysRevLett.101.151802} {\bibfield  {journal} {\bibinfo  {journal}
  {Phys. Rev. Lett.}\ }\textbf {\bibinfo {volume} {101}},\ \bibinfo {pages}
  {151802} (\bibinfo {year} {2008})},\ \Eprint {http://arxiv.org/abs/0807.1427}
  {arXiv:0807.1427 [hep-ex]} \BibitemShut {NoStop}%
\bibitem [{\citenamefont {Lees}\ \emph
  {et~al.}(2013{\natexlab{a}})\citenamefont {Lees} \emph
  {et~al.}}]{BaBar:2012wey}%
  \BibitemOpen
  \bibfield  {author} {\bibinfo {author} {\bibfnamefont {J.~P.}\ \bibnamefont
  {Lees}} \emph {et~al.} (\bibinfo {collaboration} {BaBar}),\ }\href {\doibase
  10.1103/PhysRevD.87.031102} {\bibfield  {journal} {\bibinfo  {journal} {Phys.
  Rev. D}\ }\textbf {\bibinfo {volume} {87}},\ \bibinfo {pages} {031102}
  (\bibinfo {year} {2013}{\natexlab{a}})},\ \bibinfo {note} {[Erratum:
  Phys.Rev.D 87, 059903 (2013)]},\ \Eprint {http://arxiv.org/abs/1210.0287}
  {arXiv:1210.0287 [hep-ex]} \BibitemShut {NoStop}%
\bibitem [{\citenamefont {Lees}\ \emph
  {et~al.}(2013{\natexlab{b}})\citenamefont {Lees} \emph
  {et~al.}}]{BaBar:2012sau}%
  \BibitemOpen
  \bibfield  {author} {\bibinfo {author} {\bibfnamefont {J.~P.}\ \bibnamefont
  {Lees}} \emph {et~al.} (\bibinfo {collaboration} {BaBar}),\ }\href {\doibase
  10.1103/PhysRevD.88.071102} {\bibfield  {journal} {\bibinfo  {journal} {Phys.
  Rev. D}\ }\textbf {\bibinfo {volume} {88}},\ \bibinfo {pages} {071102}
  (\bibinfo {year} {2013}{\natexlab{b}})},\ \Eprint
  {http://arxiv.org/abs/1210.5669} {arXiv:1210.5669 [hep-ex]} \BibitemShut
  {NoStop}%
\bibitem [{\citenamefont {Lees}\ \emph {et~al.}(2011)\citenamefont {Lees} \emph
  {et~al.}}]{BaBar:2011kau}%
  \BibitemOpen
  \bibfield  {author} {\bibinfo {author} {\bibfnamefont {J.~P.}\ \bibnamefont
  {Lees}} \emph {et~al.} (\bibinfo {collaboration} {BaBar}),\ }\href {\doibase
  10.1103/PhysRevLett.107.221803} {\bibfield  {journal} {\bibinfo  {journal}
  {Phys. Rev. Lett.}\ }\textbf {\bibinfo {volume} {107}},\ \bibinfo {pages}
  {221803} (\bibinfo {year} {2011})},\ \Eprint {http://arxiv.org/abs/1108.3549}
  {arXiv:1108.3549 [hep-ex]} \BibitemShut {NoStop}%
\bibitem [{\citenamefont {Hiller}(2004)}]{Hiller:2004ii}%
  \BibitemOpen
  \bibfield  {author} {\bibinfo {author} {\bibfnamefont {G.}~\bibnamefont
  {Hiller}},\ }\href {\doibase 10.1103/PhysRevD.70.034018} {\bibfield
  {journal} {\bibinfo  {journal} {Phys. Rev. D}\ }\textbf {\bibinfo {volume}
  {70}},\ \bibinfo {pages} {034018} (\bibinfo {year} {2004})},\ \Eprint
  {http://arxiv.org/abs/hep-ph/0404220} {arXiv:hep-ph/0404220} \BibitemShut
  {NoStop}%
\bibitem [{\citenamefont {Ball}\ and\ \citenamefont
  {Zwicky}(2005)}]{Ball:2004ye}%
  \BibitemOpen
  \bibfield  {author} {\bibinfo {author} {\bibfnamefont {P.}~\bibnamefont
  {Ball}}\ and\ \bibinfo {author} {\bibfnamefont {R.}~\bibnamefont {Zwicky}},\
  }\href {\doibase 10.1103/PhysRevD.71.014015} {\bibfield  {journal} {\bibinfo
  {journal} {Phys. Rev. D}\ }\textbf {\bibinfo {volume} {71}},\ \bibinfo
  {pages} {014015} (\bibinfo {year} {2005})},\ \Eprint
  {http://arxiv.org/abs/hep-ph/0406232} {arXiv:hep-ph/0406232} \BibitemShut
  {NoStop}%
\bibitem [{\citenamefont {Lutz}\ \emph {et~al.}(2013)\citenamefont {Lutz} \emph
  {et~al.}}]{Belle:2013tnz}%
  \BibitemOpen
  \bibfield  {author} {\bibinfo {author} {\bibfnamefont {O.}~\bibnamefont
  {Lutz}} \emph {et~al.} (\bibinfo {collaboration} {Belle}),\ }\href {\doibase
  10.1103/PhysRevD.87.111103} {\bibfield  {journal} {\bibinfo  {journal} {Phys.
  Rev. D}\ }\textbf {\bibinfo {volume} {87}},\ \bibinfo {pages} {111103}
  (\bibinfo {year} {2013})},\ \Eprint {http://arxiv.org/abs/1303.3719}
  {arXiv:1303.3719 [hep-ex]} \BibitemShut {NoStop}%
\bibitem [{\citenamefont {del Amo~Sanchez}\ \emph {et~al.}(2010)\citenamefont
  {del Amo~Sanchez} \emph {et~al.}}]{BaBar:2010oqg}%
  \BibitemOpen
  \bibfield  {author} {\bibinfo {author} {\bibfnamefont {P.}~\bibnamefont {del
  Amo~Sanchez}} \emph {et~al.} (\bibinfo {collaboration} {BaBar}),\ }\href
  {\doibase 10.1103/PhysRevD.82.112002} {\bibfield  {journal} {\bibinfo
  {journal} {Phys. Rev. D}\ }\textbf {\bibinfo {volume} {82}},\ \bibinfo
  {pages} {112002} (\bibinfo {year} {2010})},\ \Eprint
  {http://arxiv.org/abs/1009.1529} {arXiv:1009.1529 [hep-ex]} \BibitemShut
  {NoStop}%
\bibitem [{\citenamefont {Aaij}\ \emph {et~al.}(2015)\citenamefont {Aaij} \emph
  {et~al.}}]{LHCb:2015nkv}%
  \BibitemOpen
  \bibfield  {author} {\bibinfo {author} {\bibfnamefont {R.}~\bibnamefont
  {Aaij}} \emph {et~al.} (\bibinfo {collaboration} {LHCb}),\ }\href {\doibase
  10.1103/PhysRevLett.115.161802} {\bibfield  {journal} {\bibinfo  {journal}
  {Phys. Rev. Lett.}\ }\textbf {\bibinfo {volume} {115}},\ \bibinfo {pages}
  {161802} (\bibinfo {year} {2015})},\ \Eprint
  {http://arxiv.org/abs/1508.04094} {arXiv:1508.04094 [hep-ex]} \BibitemShut
  {NoStop}%
\bibitem [{\citenamefont {Aaij}\ \emph {et~al.}(2017)\citenamefont {Aaij} \emph
  {et~al.}}]{LHCb:2016awg}%
  \BibitemOpen
  \bibfield  {author} {\bibinfo {author} {\bibfnamefont {R.}~\bibnamefont
  {Aaij}} \emph {et~al.} (\bibinfo {collaboration} {LHCb}),\ }\href {\doibase
  10.1103/PhysRevD.95.071101} {\bibfield  {journal} {\bibinfo  {journal} {Phys.
  Rev. D}\ }\textbf {\bibinfo {volume} {95}},\ \bibinfo {pages} {071101}
  (\bibinfo {year} {2017})},\ \Eprint {http://arxiv.org/abs/1612.07818}
  {arXiv:1612.07818 [hep-ex]} \BibitemShut {NoStop}%
\bibitem [{\citenamefont {Aubert}\ \emph {et~al.}(2009)\citenamefont {Aubert}
  \emph {et~al.}}]{BaBar:2008jdv}%
  \BibitemOpen
  \bibfield  {author} {\bibinfo {author} {\bibfnamefont {B.}~\bibnamefont
  {Aubert}} \emph {et~al.} (\bibinfo {collaboration} {BaBar}),\ }\href
  {\doibase 10.1103/PhysRevLett.102.091803} {\bibfield  {journal} {\bibinfo
  {journal} {Phys. Rev. Lett.}\ }\textbf {\bibinfo {volume} {102}},\ \bibinfo
  {pages} {091803} (\bibinfo {year} {2009})},\ \Eprint
  {http://arxiv.org/abs/0807.4119} {arXiv:0807.4119 [hep-ex]} \BibitemShut
  {NoStop}%
\bibitem [{\citenamefont {Wei}\ \emph {et~al.}(2009)\citenamefont {Wei} \emph
  {et~al.}}]{Belle:2009zue}%
  \BibitemOpen
  \bibfield  {author} {\bibinfo {author} {\bibfnamefont {J.~T.}\ \bibnamefont
  {Wei}} \emph {et~al.} (\bibinfo {collaboration} {Belle}),\ }\href {\doibase
  10.1103/PhysRevLett.103.171801} {\bibfield  {journal} {\bibinfo  {journal}
  {Phys. Rev. Lett.}\ }\textbf {\bibinfo {volume} {103}},\ \bibinfo {pages}
  {171801} (\bibinfo {year} {2009})},\ \Eprint {http://arxiv.org/abs/0904.0770}
  {arXiv:0904.0770 [hep-ex]} \BibitemShut {NoStop}%
\bibitem [{\citenamefont {Deshpande}\ \emph {et~al.}(2006)\citenamefont
  {Deshpande}, \citenamefont {Eilam},\ and\ \citenamefont
  {Jiang}}]{Deshpande:2005mb}%
  \BibitemOpen
  \bibfield  {author} {\bibinfo {author} {\bibfnamefont {N.~G.}\ \bibnamefont
  {Deshpande}}, \bibinfo {author} {\bibfnamefont {G.}~\bibnamefont {Eilam}}, \
  and\ \bibinfo {author} {\bibfnamefont {J.}~\bibnamefont {Jiang}},\ }\href
  {\doibase 10.1016/j.physletb.2005.10.050} {\bibfield  {journal} {\bibinfo
  {journal} {Phys. Lett. B}\ }\textbf {\bibinfo {volume} {632}},\ \bibinfo
  {pages} {212} (\bibinfo {year} {2006})},\ \Eprint
  {http://arxiv.org/abs/hep-ph/0509081} {arXiv:hep-ph/0509081} \BibitemShut
  {NoStop}%
\bibitem [{\citenamefont {Marciano}\ and\ \citenamefont
  {Parsa}(1996)}]{Marciano:1996wy}%
  \BibitemOpen
  \bibfield  {author} {\bibinfo {author} {\bibfnamefont {W.~J.}\ \bibnamefont
  {Marciano}}\ and\ \bibinfo {author} {\bibfnamefont {Z.}~\bibnamefont
  {Parsa}},\ }\href {\doibase 10.1103/PhysRevD.53.R1} {\bibfield  {journal}
  {\bibinfo  {journal} {Phys. Rev. D}\ }\textbf {\bibinfo {volume} {53}},\
  \bibinfo {pages} {R1} (\bibinfo {year} {1996})}\BibitemShut {NoStop}%
\bibitem [{\citenamefont {Cortina~Gil}\ \emph
  {et~al.}(2021{\natexlab{a}})\citenamefont {Cortina~Gil} \emph
  {et~al.}}]{NA62:2020xlg}%
  \BibitemOpen
  \bibfield  {author} {\bibinfo {author} {\bibfnamefont {E.}~\bibnamefont
  {Cortina~Gil}} \emph {et~al.} (\bibinfo {collaboration} {NA62}),\ }\href
  {\doibase 10.1007/JHEP03(2021)058} {\bibfield  {journal} {\bibinfo  {journal}
  {JHEP}\ }\textbf {\bibinfo {volume} {03}},\ \bibinfo {pages} {058} (\bibinfo
  {year} {2021}{\natexlab{a}})},\ \Eprint {http://arxiv.org/abs/2011.11329}
  {arXiv:2011.11329 [hep-ex]} \BibitemShut {NoStop}%
\bibitem [{\citenamefont {Cortina~Gil}\ \emph
  {et~al.}(2021{\natexlab{b}})\citenamefont {Cortina~Gil} \emph
  {et~al.}}]{NA62:2021zjw}%
  \BibitemOpen
  \bibfield  {author} {\bibinfo {author} {\bibfnamefont {E.}~\bibnamefont
  {Cortina~Gil}} \emph {et~al.} (\bibinfo {collaboration} {NA62}),\ }\href
  {\doibase 10.1007/JHEP06(2021)093} {\bibfield  {journal} {\bibinfo  {journal}
  {JHEP}\ }\textbf {\bibinfo {volume} {06}},\ \bibinfo {pages} {093} (\bibinfo
  {year} {2021}{\natexlab{b}})},\ \Eprint {http://arxiv.org/abs/2103.15389}
  {arXiv:2103.15389 [hep-ex]} \BibitemShut {NoStop}%
\bibitem [{\citenamefont {Cortina~Gil}\ \emph
  {et~al.}(2021{\natexlab{c}})\citenamefont {Cortina~Gil} \emph
  {et~al.}}]{NA62:2020pwi}%
  \BibitemOpen
  \bibfield  {author} {\bibinfo {author} {\bibfnamefont {E.}~\bibnamefont
  {Cortina~Gil}} \emph {et~al.} (\bibinfo {collaboration} {NA62}),\ }\href
  {\doibase 10.1007/JHEP02(2021)201} {\bibfield  {journal} {\bibinfo  {journal}
  {JHEP}\ }\textbf {\bibinfo {volume} {02}},\ \bibinfo {pages} {201} (\bibinfo
  {year} {2021}{\natexlab{c}})},\ \Eprint {http://arxiv.org/abs/2010.07644}
  {arXiv:2010.07644 [hep-ex]} \BibitemShut {NoStop}%
\bibitem [{\citenamefont {Buras}\ \emph {et~al.}(2015)\citenamefont {Buras},
  \citenamefont {Buttazzo}, \citenamefont {Girrbach-Noe},\ and\ \citenamefont
  {Knegjens}}]{Buras:2015qea}%
  \BibitemOpen
  \bibfield  {author} {\bibinfo {author} {\bibfnamefont {A.~J.}\ \bibnamefont
  {Buras}}, \bibinfo {author} {\bibfnamefont {D.}~\bibnamefont {Buttazzo}},
  \bibinfo {author} {\bibfnamefont {J.}~\bibnamefont {Girrbach-Noe}}, \ and\
  \bibinfo {author} {\bibfnamefont {R.}~\bibnamefont {Knegjens}},\ }\href
  {\doibase 10.1007/JHEP11(2015)033} {\bibfield  {journal} {\bibinfo  {journal}
  {JHEP}\ }\textbf {\bibinfo {volume} {11}},\ \bibinfo {pages} {033} (\bibinfo
  {year} {2015})},\ \Eprint {http://arxiv.org/abs/1503.02693} {arXiv:1503.02693
  [hep-ph]} \BibitemShut {NoStop}%
\bibitem [{\citenamefont {Alavi-Harati}\ \emph {et~al.}(2004)\citenamefont
  {Alavi-Harati} \emph {et~al.}}]{KTeV:2003sls}%
  \BibitemOpen
  \bibfield  {author} {\bibinfo {author} {\bibfnamefont {A.}~\bibnamefont
  {Alavi-Harati}} \emph {et~al.} (\bibinfo {collaboration} {KTeV}),\ }\href
  {\doibase 10.1103/PhysRevLett.93.021805} {\bibfield  {journal} {\bibinfo
  {journal} {Phys. Rev. Lett.}\ }\textbf {\bibinfo {volume} {93}},\ \bibinfo
  {pages} {021805} (\bibinfo {year} {2004})},\ \Eprint
  {http://arxiv.org/abs/hep-ex/0309072} {arXiv:hep-ex/0309072} \BibitemShut
  {NoStop}%
\bibitem [{\citenamefont {Abouzaid}\ \emph {et~al.}(2008)\citenamefont
  {Abouzaid} \emph {et~al.}}]{KTeV:2008nqz}%
  \BibitemOpen
  \bibfield  {author} {\bibinfo {author} {\bibfnamefont {E.}~\bibnamefont
  {Abouzaid}} \emph {et~al.} (\bibinfo {collaboration} {KTeV}),\ }\href
  {\doibase 10.1103/PhysRevD.77.112004} {\bibfield  {journal} {\bibinfo
  {journal} {Phys. Rev. D}\ }\textbf {\bibinfo {volume} {77}},\ \bibinfo
  {pages} {112004} (\bibinfo {year} {2008})},\ \Eprint
  {http://arxiv.org/abs/0805.0031} {arXiv:0805.0031 [hep-ex]} \BibitemShut
  {NoStop}%
\bibitem [{\citenamefont {Bergsma}\ \emph {et~al.}(1985)\citenamefont {Bergsma}
  \emph {et~al.}}]{CHARM:1985anb}%
  \BibitemOpen
  \bibfield  {author} {\bibinfo {author} {\bibfnamefont {F.}~\bibnamefont
  {Bergsma}} \emph {et~al.} (\bibinfo {collaboration} {CHARM}),\ }\href
  {\doibase 10.1016/0370-2693(85)90400-9} {\bibfield  {journal} {\bibinfo
  {journal} {Phys. Lett. B}\ }\textbf {\bibinfo {volume} {157}},\ \bibinfo
  {pages} {458} (\bibinfo {year} {1985})}\BibitemShut {NoStop}%
\bibitem [{\citenamefont {Clarke}\ \emph {et~al.}(2014)\citenamefont {Clarke},
  \citenamefont {Foot},\ and\ \citenamefont {Volkas}}]{Clarke:2013aya}%
  \BibitemOpen
  \bibfield  {author} {\bibinfo {author} {\bibfnamefont {J.~D.}\ \bibnamefont
  {Clarke}}, \bibinfo {author} {\bibfnamefont {R.}~\bibnamefont {Foot}}, \ and\
  \bibinfo {author} {\bibfnamefont {R.~R.}\ \bibnamefont {Volkas}},\ }\href
  {\doibase 10.1007/JHEP02(2014)123} {\bibfield  {journal} {\bibinfo  {journal}
  {JHEP}\ }\textbf {\bibinfo {volume} {02}},\ \bibinfo {pages} {123} (\bibinfo
  {year} {2014})},\ \Eprint {http://arxiv.org/abs/1310.8042} {arXiv:1310.8042
  [hep-ph]} \BibitemShut {NoStop}%
\bibitem [{\citenamefont {Bezrukov}\ and\ \citenamefont
  {Gorbunov}(2010)}]{Bezrukov:2009yw}%
  \BibitemOpen
  \bibfield  {author} {\bibinfo {author} {\bibfnamefont {F.}~\bibnamefont
  {Bezrukov}}\ and\ \bibinfo {author} {\bibfnamefont {D.}~\bibnamefont
  {Gorbunov}},\ }\href {\doibase 10.1007/JHEP05(2010)010} {\bibfield  {journal}
  {\bibinfo  {journal} {JHEP}\ }\textbf {\bibinfo {volume} {05}},\ \bibinfo
  {pages} {010} (\bibinfo {year} {2010})},\ \Eprint
  {http://arxiv.org/abs/0912.0390} {arXiv:0912.0390 [hep-ph]} \BibitemShut
  {NoStop}%
\bibitem [{\citenamefont {Schmidt-Hoberg}\ \emph {et~al.}(2013)\citenamefont
  {Schmidt-Hoberg}, \citenamefont {Staub},\ and\ \citenamefont
  {Winkler}}]{Schmidt-Hoberg:2013hba}%
  \BibitemOpen
  \bibfield  {author} {\bibinfo {author} {\bibfnamefont {K.}~\bibnamefont
  {Schmidt-Hoberg}}, \bibinfo {author} {\bibfnamefont {F.}~\bibnamefont
  {Staub}}, \ and\ \bibinfo {author} {\bibfnamefont {M.~W.}\ \bibnamefont
  {Winkler}},\ }\href {\doibase 10.1016/j.physletb.2013.11.015} {\bibfield
  {journal} {\bibinfo  {journal} {Phys. Lett. B}\ }\textbf {\bibinfo {volume}
  {727}},\ \bibinfo {pages} {506} (\bibinfo {year} {2013})},\ \Eprint
  {http://arxiv.org/abs/1310.6752} {arXiv:1310.6752 [hep-ph]} \BibitemShut
  {NoStop}%
\bibitem [{\citenamefont {Alekhin}\ \emph {et~al.}(2016)\citenamefont {Alekhin}
  \emph {et~al.}}]{Alekhin:2015byh}%
  \BibitemOpen
  \bibfield  {author} {\bibinfo {author} {\bibfnamefont {S.}~\bibnamefont
  {Alekhin}} \emph {et~al.},\ }\href {\doibase 10.1088/0034-4885/79/12/124201}
  {\bibfield  {journal} {\bibinfo  {journal} {Rept. Prog. Phys.}\ }\textbf
  {\bibinfo {volume} {79}},\ \bibinfo {pages} {124201} (\bibinfo {year}
  {2016})},\ \Eprint {http://arxiv.org/abs/1504.04855} {arXiv:1504.04855
  [hep-ph]} \BibitemShut {NoStop}%
\bibitem [{\citenamefont {Gorbunov}\ \emph {et~al.}(2021)\citenamefont
  {Gorbunov}, \citenamefont {Krasnov},\ and\ \citenamefont
  {Suvorov}}]{Gorbunov:2021ccu}%
  \BibitemOpen
  \bibfield  {author} {\bibinfo {author} {\bibfnamefont {D.}~\bibnamefont
  {Gorbunov}}, \bibinfo {author} {\bibfnamefont {I.}~\bibnamefont {Krasnov}}, \
  and\ \bibinfo {author} {\bibfnamefont {S.}~\bibnamefont {Suvorov}},\ }\href
  {\doibase 10.1016/j.physletb.2021.136524} {\bibfield  {journal} {\bibinfo
  {journal} {Phys. Lett. B}\ }\textbf {\bibinfo {volume} {820}},\ \bibinfo
  {pages} {136524} (\bibinfo {year} {2021})},\ \Eprint
  {http://arxiv.org/abs/2105.11102} {arXiv:2105.11102 [hep-ph]} \BibitemShut
  {NoStop}%
\bibitem [{\citenamefont {Shifman}\ \emph {et~al.}(1978)\citenamefont
  {Shifman}, \citenamefont {Vainshtein},\ and\ \citenamefont
  {Zakharov}}]{Shifman:1978zn}%
  \BibitemOpen
  \bibfield  {author} {\bibinfo {author} {\bibfnamefont {M.~A.}\ \bibnamefont
  {Shifman}}, \bibinfo {author} {\bibfnamefont {A.~I.}\ \bibnamefont
  {Vainshtein}}, \ and\ \bibinfo {author} {\bibfnamefont {V.~I.}\ \bibnamefont
  {Zakharov}},\ }\href {\doibase 10.1016/0370-2693(78)90481-1} {\bibfield
  {journal} {\bibinfo  {journal} {Phys. Lett. B}\ }\textbf {\bibinfo {volume}
  {78}},\ \bibinfo {pages} {443} (\bibinfo {year} {1978})}\BibitemShut
  {NoStop}%
\bibitem [{\citenamefont {B\'elanger}\ \emph {et~al.}(2018)\citenamefont
  {B\'elanger}, \citenamefont {Boudjema}, \citenamefont {Goudelis},
  \citenamefont {Pukhov},\ and\ \citenamefont {Zaldivar}}]{Belanger:2018ccd}%
  \BibitemOpen
  \bibfield  {author} {\bibinfo {author} {\bibfnamefont {G.}~\bibnamefont
  {B\'elanger}}, \bibinfo {author} {\bibfnamefont {F.}~\bibnamefont
  {Boudjema}}, \bibinfo {author} {\bibfnamefont {A.}~\bibnamefont {Goudelis}},
  \bibinfo {author} {\bibfnamefont {A.}~\bibnamefont {Pukhov}}, \ and\ \bibinfo
  {author} {\bibfnamefont {B.}~\bibnamefont {Zaldivar}},\ }\href {\doibase
  10.1016/j.cpc.2018.04.027} {\bibfield  {journal} {\bibinfo  {journal}
  {Comput. Phys. Commun.}\ }\textbf {\bibinfo {volume} {231}},\ \bibinfo
  {pages} {173} (\bibinfo {year} {2018})},\ \Eprint
  {http://arxiv.org/abs/1801.03509} {arXiv:1801.03509 [hep-ph]} \BibitemShut
  {NoStop}%
\bibitem [{\citenamefont {Abdelhameed}\ \emph {et~al.}(2019)\citenamefont
  {Abdelhameed} \emph {et~al.}}]{CRESST:2019jnq}%
  \BibitemOpen
  \bibfield  {author} {\bibinfo {author} {\bibfnamefont {A.~H.}\ \bibnamefont
  {Abdelhameed}} \emph {et~al.} (\bibinfo {collaboration} {CRESST}),\ }\href
  {\doibase 10.1103/PhysRevD.100.102002} {\bibfield  {journal} {\bibinfo
  {journal} {Phys. Rev. D}\ }\textbf {\bibinfo {volume} {100}},\ \bibinfo
  {pages} {102002} (\bibinfo {year} {2019})},\ \Eprint
  {http://arxiv.org/abs/1904.00498} {arXiv:1904.00498 [astro-ph.CO]}
  \BibitemShut {NoStop}%
\bibitem [{\citenamefont {Agnes}\ \emph {et~al.}(2018)\citenamefont {Agnes}
  \emph {et~al.}}]{DarkSide:2018bpj}%
  \BibitemOpen
  \bibfield  {author} {\bibinfo {author} {\bibfnamefont {P.}~\bibnamefont
  {Agnes}} \emph {et~al.} (\bibinfo {collaboration} {DarkSide}),\ }\href
  {\doibase 10.1103/PhysRevLett.121.081307} {\bibfield  {journal} {\bibinfo
  {journal} {Phys. Rev. Lett.}\ }\textbf {\bibinfo {volume} {121}},\ \bibinfo
  {pages} {081307} (\bibinfo {year} {2018})},\ \Eprint
  {http://arxiv.org/abs/1802.06994} {arXiv:1802.06994 [astro-ph.HE]}
  \BibitemShut {NoStop}%
\bibitem [{\citenamefont {Padmanabhan}\ and\ \citenamefont
  {Finkbeiner}(2005)}]{Padmanabhan:2005es}%
  \BibitemOpen
  \bibfield  {author} {\bibinfo {author} {\bibfnamefont {N.}~\bibnamefont
  {Padmanabhan}}\ and\ \bibinfo {author} {\bibfnamefont {D.~P.}\ \bibnamefont
  {Finkbeiner}},\ }\href {\doibase 10.1103/PhysRevD.72.023508} {\bibfield
  {journal} {\bibinfo  {journal} {Phys. Rev. D}\ }\textbf {\bibinfo {volume}
  {72}},\ \bibinfo {pages} {023508} (\bibinfo {year} {2005})},\ \Eprint
  {http://arxiv.org/abs/astro-ph/0503486} {arXiv:astro-ph/0503486} \BibitemShut
  {NoStop}%
\bibitem [{\citenamefont {Cyburt}\ \emph {et~al.}(2016)\citenamefont {Cyburt},
  \citenamefont {Fields}, \citenamefont {Olive},\ and\ \citenamefont
  {Yeh}}]{Cyburt:2015mya}%
  \BibitemOpen
  \bibfield  {author} {\bibinfo {author} {\bibfnamefont {R.~H.}\ \bibnamefont
  {Cyburt}}, \bibinfo {author} {\bibfnamefont {B.~D.}\ \bibnamefont {Fields}},
  \bibinfo {author} {\bibfnamefont {K.~A.}\ \bibnamefont {Olive}}, \ and\
  \bibinfo {author} {\bibfnamefont {T.-H.}\ \bibnamefont {Yeh}},\ }\href
  {\doibase 10.1103/RevModPhys.88.015004} {\bibfield  {journal} {\bibinfo
  {journal} {Rev. Mod. Phys.}\ }\textbf {\bibinfo {volume} {88}},\ \bibinfo
  {pages} {015004} (\bibinfo {year} {2016})},\ \Eprint
  {http://arxiv.org/abs/1505.01076} {arXiv:1505.01076 [astro-ph.CO]}
  \BibitemShut {NoStop}%
\bibitem [{\citenamefont {Fradette}\ and\ \citenamefont
  {Pospelov}(2017)}]{Fradette:2017sdd}%
  \BibitemOpen
  \bibfield  {author} {\bibinfo {author} {\bibfnamefont {A.}~\bibnamefont
  {Fradette}}\ and\ \bibinfo {author} {\bibfnamefont {M.}~\bibnamefont
  {Pospelov}},\ }\href {\doibase 10.1103/PhysRevD.96.075033} {\bibfield
  {journal} {\bibinfo  {journal} {Phys. Rev. D}\ }\textbf {\bibinfo {volume}
  {96}},\ \bibinfo {pages} {075033} (\bibinfo {year} {2017})},\ \Eprint
  {http://arxiv.org/abs/1706.01920} {arXiv:1706.01920 [hep-ph]} \BibitemShut
  {NoStop}%
\bibitem [{\citenamefont {Kohri}(2001)}]{Kohri:2001jx}%
  \BibitemOpen
  \bibfield  {author} {\bibinfo {author} {\bibfnamefont {K.}~\bibnamefont
  {Kohri}},\ }\href {\doibase 10.1103/PhysRevD.64.043515} {\bibfield  {journal}
  {\bibinfo  {journal} {Phys. Rev. D}\ }\textbf {\bibinfo {volume} {64}},\
  \bibinfo {pages} {043515} (\bibinfo {year} {2001})},\ \Eprint
  {http://arxiv.org/abs/astro-ph/0103411} {arXiv:astro-ph/0103411} \BibitemShut
  {NoStop}%
\bibitem [{\citenamefont {Hirata}\ \emph {et~al.}(1987)\citenamefont {Hirata}
  \emph {et~al.}}]{Kamiokande-II:1987idp}%
  \BibitemOpen
  \bibfield  {author} {\bibinfo {author} {\bibfnamefont {K.}~\bibnamefont
  {Hirata}} \emph {et~al.} (\bibinfo {collaboration} {Kamiokande-II}),\ }\href
  {\doibase 10.1103/PhysRevLett.58.1490} {\bibfield  {journal} {\bibinfo
  {journal} {Phys. Rev. Lett.}\ }\textbf {\bibinfo {volume} {58}},\ \bibinfo
  {pages} {1490} (\bibinfo {year} {1987})}\BibitemShut {NoStop}%
\bibitem [{\citenamefont {Bionta}\ \emph {et~al.}(1987)\citenamefont {Bionta}
  \emph {et~al.}}]{Bionta:1987qt}%
  \BibitemOpen
  \bibfield  {author} {\bibinfo {author} {\bibfnamefont {R.~M.}\ \bibnamefont
  {Bionta}} \emph {et~al.},\ }\href {\doibase 10.1103/PhysRevLett.58.1494}
  {\bibfield  {journal} {\bibinfo  {journal} {Phys. Rev. Lett.}\ }\textbf
  {\bibinfo {volume} {58}},\ \bibinfo {pages} {1494} (\bibinfo {year}
  {1987})}\BibitemShut {NoStop}%
\bibitem [{\citenamefont {Raffelt}\ and\ \citenamefont
  {Seckel}(1988)}]{Raffelt:1987yt}%
  \BibitemOpen
  \bibfield  {author} {\bibinfo {author} {\bibfnamefont {G.}~\bibnamefont
  {Raffelt}}\ and\ \bibinfo {author} {\bibfnamefont {D.}~\bibnamefont
  {Seckel}},\ }\href {\doibase 10.1103/PhysRevLett.60.1793} {\bibfield
  {journal} {\bibinfo  {journal} {Phys. Rev. Lett.}\ }\textbf {\bibinfo
  {volume} {60}},\ \bibinfo {pages} {1793} (\bibinfo {year}
  {1988})}\BibitemShut {NoStop}%
\bibitem [{\citenamefont {Turner}(1988)}]{Turner:1987by}%
  \BibitemOpen
  \bibfield  {author} {\bibinfo {author} {\bibfnamefont {M.~S.}\ \bibnamefont
  {Turner}},\ }\href {\doibase 10.1103/PhysRevLett.60.1797} {\bibfield
  {journal} {\bibinfo  {journal} {Phys. Rev. Lett.}\ }\textbf {\bibinfo
  {volume} {60}},\ \bibinfo {pages} {1797} (\bibinfo {year}
  {1988})}\BibitemShut {NoStop}%
\bibitem [{\citenamefont {Dicus}\ \emph {et~al.}(1978)\citenamefont {Dicus},
  \citenamefont {Kolb}, \citenamefont {Teplitz},\ and\ \citenamefont
  {Wagoner}}]{Dicus:1978fp}%
  \BibitemOpen
  \bibfield  {author} {\bibinfo {author} {\bibfnamefont {D.~A.}\ \bibnamefont
  {Dicus}}, \bibinfo {author} {\bibfnamefont {E.~W.}\ \bibnamefont {Kolb}},
  \bibinfo {author} {\bibfnamefont {V.~L.}\ \bibnamefont {Teplitz}}, \ and\
  \bibinfo {author} {\bibfnamefont {R.~V.}\ \bibnamefont {Wagoner}},\ }\href
  {\doibase 10.1103/PhysRevD.18.1829} {\bibfield  {journal} {\bibinfo
  {journal} {Phys. Rev. D}\ }\textbf {\bibinfo {volume} {18}},\ \bibinfo
  {pages} {1829} (\bibinfo {year} {1978})}\BibitemShut {NoStop}%
\bibitem [{\citenamefont {Dev}\ \emph {et~al.}(2020)\citenamefont {Dev},
  \citenamefont {Mohapatra},\ and\ \citenamefont {Zhang}}]{Dev:2020eam}%
  \BibitemOpen
  \bibfield  {author} {\bibinfo {author} {\bibfnamefont {P.~S.~B.}\
  \bibnamefont {Dev}}, \bibinfo {author} {\bibfnamefont {R.~N.}\ \bibnamefont
  {Mohapatra}}, \ and\ \bibinfo {author} {\bibfnamefont {Y.}~\bibnamefont
  {Zhang}},\ }\href {\doibase 10.1088/1475-7516/2020/08/003} {\bibfield
  {journal} {\bibinfo  {journal} {JCAP}\ }\textbf {\bibinfo {volume} {08}},\
  \bibinfo {pages} {003} (\bibinfo {year} {2020})},\ \bibinfo {note} {[Erratum:
  JCAP 11, E01 (2020)]},\ \Eprint {http://arxiv.org/abs/2005.00490}
  {arXiv:2005.00490 [hep-ph]} \BibitemShut {NoStop}%
\bibitem [{\citenamefont {Cepeda}\ \emph {et~al.}(2019)\citenamefont {Cepeda}
  \emph {et~al.}}]{Cepeda:2019klc}%
  \BibitemOpen
  \bibfield  {author} {\bibinfo {author} {\bibfnamefont {M.}~\bibnamefont
  {Cepeda}} \emph {et~al.},\ }\href {\doibase 10.23731/CYRM-2019-007.221}
  {\bibfield  {journal} {\bibinfo  {journal} {CERN Yellow Rep. Monogr.}\
  }\textbf {\bibinfo {volume} {7}},\ \bibinfo {pages} {221} (\bibinfo {year}
  {2019})},\ \Eprint {http://arxiv.org/abs/1902.00134} {arXiv:1902.00134
  [hep-ph]} \BibitemShut {NoStop}%
\bibitem [{\citenamefont {Baer}\ \emph {et~al.}(2013)\citenamefont {Baer} \emph
  {et~al.}}]{Baer:2013cma}%
  \BibitemOpen
  \bibfield  {author} {\bibinfo {author} {\bibfnamefont {H.}~\bibnamefont
  {Baer}} \emph {et~al.},\ }\href@noop {} {\bibfield  {journal} {\bibinfo
  {journal} {The International Linear Collider Technical Design Report - Volume
  2: Physics}\ } (\bibinfo {year} {2013})},\ \Eprint
  {http://arxiv.org/abs/1306.6352} {arXiv:1306.6352 [hep-ph]} \BibitemShut
  {NoStop}%
\bibitem [{\citenamefont {Barklow}\ \emph {et~al.}(2018)\citenamefont
  {Barklow}, \citenamefont {Fujii}, \citenamefont {Jung}, \citenamefont {Karl},
  \citenamefont {List}, \citenamefont {Ogawa}, \citenamefont {Peskin},\ and\
  \citenamefont {Tian}}]{Barklow:2017suo}%
  \BibitemOpen
  \bibfield  {author} {\bibinfo {author} {\bibfnamefont {T.}~\bibnamefont
  {Barklow}}, \bibinfo {author} {\bibfnamefont {K.}~\bibnamefont {Fujii}},
  \bibinfo {author} {\bibfnamefont {S.}~\bibnamefont {Jung}}, \bibinfo {author}
  {\bibfnamefont {R.}~\bibnamefont {Karl}}, \bibinfo {author} {\bibfnamefont
  {J.}~\bibnamefont {List}}, \bibinfo {author} {\bibfnamefont {T.}~\bibnamefont
  {Ogawa}}, \bibinfo {author} {\bibfnamefont {M.~E.}\ \bibnamefont {Peskin}}, \
  and\ \bibinfo {author} {\bibfnamefont {J.}~\bibnamefont {Tian}},\ }\href
  {\doibase 10.1103/PhysRevD.97.053003} {\bibfield  {journal} {\bibinfo
  {journal} {Phys. Rev. D}\ }\textbf {\bibinfo {volume} {97}},\ \bibinfo
  {pages} {053003} (\bibinfo {year} {2018})},\ \Eprint
  {http://arxiv.org/abs/1708.08912} {arXiv:1708.08912 [hep-ph]} \BibitemShut
  {NoStop}%
\bibitem [{\citenamefont {Aushev}\ \emph {et~al.}(2010)\citenamefont {Aushev}
  \emph {et~al.}}]{Aushev:2010bq}%
  \BibitemOpen
  \bibfield  {author} {\bibinfo {author} {\bibfnamefont {T.}~\bibnamefont
  {Aushev}} \emph {et~al.},\ }\href@noop {} {\  (\bibinfo {year} {2010})},\
  \Eprint {http://arxiv.org/abs/1002.5012} {arXiv:1002.5012 [hep-ex]}
  \BibitemShut {NoStop}%
\bibitem [{\citenamefont {Ahn}\ \emph {et~al.}(2019)\citenamefont {Ahn} \emph
  {et~al.}}]{KOTO:2018dsc}%
  \BibitemOpen
  \bibfield  {author} {\bibinfo {author} {\bibfnamefont {J.~K.}\ \bibnamefont
  {Ahn}} \emph {et~al.} (\bibinfo {collaboration} {KOTO}),\ }\href {\doibase
  10.1103/PhysRevLett.122.021802} {\bibfield  {journal} {\bibinfo  {journal}
  {Phys. Rev. Lett.}\ }\textbf {\bibinfo {volume} {122}},\ \bibinfo {pages}
  {021802} (\bibinfo {year} {2019})},\ \Eprint
  {http://arxiv.org/abs/1810.09655} {arXiv:1810.09655 [hep-ex]} \BibitemShut
  {NoStop}%
\bibitem [{\citenamefont {Akerib}\ \emph {et~al.}(2015)\citenamefont {Akerib}
  \emph {et~al.}}]{LZ:2015kxe}%
  \BibitemOpen
  \bibfield  {author} {\bibinfo {author} {\bibfnamefont {D.~S.}\ \bibnamefont
  {Akerib}} \emph {et~al.} (\bibinfo {collaboration} {LZ}),\ }\href@noop {} {\
  (\bibinfo {year} {2015})},\ \Eprint {http://arxiv.org/abs/1509.02910}
  {arXiv:1509.02910 [physics.ins-det]} \BibitemShut {NoStop}%
\bibitem [{\citenamefont {Cushman}\ \emph {et~al.}(2013)\citenamefont {Cushman}
  \emph {et~al.}}]{Cushman:2013zza}%
  \BibitemOpen
  \bibfield  {author} {\bibinfo {author} {\bibfnamefont {P.}~\bibnamefont
  {Cushman}} \emph {et~al.},\ }in\ \href@noop {} {\emph {\bibinfo {booktitle}
  {{Community Summer Study 2013}: {Snowmass on the Mississippi}}}}\ (\bibinfo
  {year} {2013})\ \Eprint {http://arxiv.org/abs/1310.8327} {arXiv:1310.8327
  [hep-ex]} \BibitemShut {NoStop}%
\bibitem [{new()}]{news}%
  \BibitemOpen
  \href@noop {} {\bibinfo  {journal} {NEWS-G collaboration https://news-g.org}\
  }\BibitemShut {NoStop}%
\bibitem [{\citenamefont {Kanemura}\ \emph {et~al.}(2015)\citenamefont
  {Kanemura}, \citenamefont {Moroi},\ and\ \citenamefont
  {Tanabe}}]{Kanemura:2015cxa}%
  \BibitemOpen
\bibfield  {journal} {  }\bibfield  {author} {\bibinfo {author} {\bibfnamefont
  {S.}~\bibnamefont {Kanemura}}, \bibinfo {author} {\bibfnamefont
  {T.}~\bibnamefont {Moroi}}, \ and\ \bibinfo {author} {\bibfnamefont
  {T.}~\bibnamefont {Tanabe}},\ }\href {\doibase
  10.1016/j.physletb.2015.10.002} {\bibfield  {journal} {\bibinfo  {journal}
  {Phys. Lett. B}\ }\textbf {\bibinfo {volume} {751}},\ \bibinfo {pages} {25}
  (\bibinfo {year} {2015})},\ \Eprint {http://arxiv.org/abs/1507.02809}
  {arXiv:1507.02809 [hep-ph]} \BibitemShut {NoStop}%
\bibitem [{\citenamefont {Sakaki}\ and\ \citenamefont
  {Ueda}(2021)}]{Sakaki:2020mqb}%
  \BibitemOpen
  \bibfield  {author} {\bibinfo {author} {\bibfnamefont {Y.}~\bibnamefont
  {Sakaki}}\ and\ \bibinfo {author} {\bibfnamefont {D.}~\bibnamefont {Ueda}},\
  }\href {\doibase 10.1103/PhysRevD.103.035024} {\bibfield  {journal} {\bibinfo
   {journal} {Phys. Rev. D}\ }\textbf {\bibinfo {volume} {103}},\ \bibinfo
  {pages} {035024} (\bibinfo {year} {2021})},\ \Eprint
  {http://arxiv.org/abs/2009.13790} {arXiv:2009.13790 [hep-ph]} \BibitemShut
  {NoStop}%
\end{thebibliography}%

\end{document}